\documentclass[a4paper, 12pt]{article}

\usepackage{amssymb,amsthm, amsmath, amscd, amsfonts,multicol}
\usepackage{pgf,epsfig,multicol,graphics,pdfsync,url}
\usepackage{t1enc}
\usepackage[latin1]{inputenc}
\usepackage[english]{babel}
\usepackage{textcomp}
\usepackage{floatrow}

\newcommand{\ro}{\mathrm}

\newcommand{\D}{\ro{d}}
\newcommand{\eps}{\varepsilon}

\newcommand{\epsi}{\varepsilon}
\newcommand{\R}{{\mathbb{R}}}
\newcommand{\C}{{\mathbb{C}}}
\newcommand{\N}{{\mathbb{N}}}
\newcommand{\Z}{{\mathbb{Z}}}
\newcommand{\Q}{{\mathbb{Q}}}
\newcommand{\T}{{\mathbb{T}}}
\newcommand{\E}{{\mathrm{e}}}
\newcommand{\BF}{\mathcal{F}}

\newcommand{\Hi}{{\mathcal{H}}}

\newcommand{\Or}{{\mathcal{O}}}
\newcommand{\bu}{\Xi}

\newcommand{\I}{\ro{i}}

\newcommand{\la}{\langle}

\renewcommand{\tilde}{\widetilde}

\theoremstyle{plain}
\newtheorem{satz}{satz}[section]
\newtheorem{thm}[satz]{Theorem}
\newtheorem{lem}[satz]{Lemma}
\newtheorem*{lem*}{Lemma}
\newtheorem{cor}[satz]{Corollary}
\newtheorem{defi}[satz]{Definition}
\newtheorem{prop}[satz]{Proposition}
\newtheorem{ass}{Assumption}

\theoremstyle{definition}

\newtheorem*{thm*}{Theorem}

\newenvironment{pro}[1][]{\noindent{\it Proof#1.\ }}{\hfill $\Box$}

\title{Peierls substitution   for magnetic Bloch bands\footnote{This work was supported by the German Science Foundation within the SFB TR 71.}}
\author{Silvia Freund and Stefan Teufel}

\date{\today}

\begin{document}

\maketitle
\thispagestyle{empty}

\vspace{-9mm}
\begin{center}
		  Eberhard Karls Universit\"at T\"ubingen, Mathematisches Institut, \linebreak
	Auf der Morgenstelle 10, 72076 T\"ubingen, Germany. \linebreak
	{\footnotesize \texttt{stefan.teufel@uni-tuebingen.de}}\\
	{\footnotesize \texttt{sima@fa.uni-tuebingen.de}}
\end{center}
\begin{abstract}
We consider the one-particle Schr\"odinger operator in two dimensions  with a periodic potential and a strong constant magnetic field perturbed by slowly varying non-periodic scalar  and vector potentials, $\phi(\epsi x)$ and $A(\epsi x)$, for $\epsi\ll 1$. For each isolated family of 
magnetic Bloch bands we derive an effective Hamiltonian that is unitarily equivalent to the restriction of the Schr\"odinger operator to a corresponding almost invariant    subspace. At 
 leading order, our effective Hamiltonian  can be interpreted as the Peierls substitution Hamiltonian widely used in  physics for non-magnetic Bloch bands. However, while for non-magnetic Bloch bands the corresponding result is well understood, both on a heuristic and on a rigorous level, for magnetic Bloch bands it is not  clear how to  even define a Peierls substitution Hamiltonian beyond a formal expression.  The source of the  difficulty is 
 a topological obstruction:
 In contrast to the non-magnetic case,  magnetic Bloch bundles are generically not trivializable.
 As a consequence,    Peierls substitution Hamiltonians  for magnetic Bloch bands turn out to be pseudodifferential operators acting on sections of non-trivial vector bundles over a two-torus, the reduced Brillouin zone.
Part of our contribution is the construction of a suitable Weyl calculus for such pseudo\-differential operators. 

As an application of our results we construct a new family of canonical  one-band Hamiltonians $H^B_{\theta,q}$   for magnetic Bloch bands with Chern number $\theta\in \Z$ that     generalizes  the Hofstadter model  $H^B_{\rm Hof} = H^B_{0,1}$  for a single non-magnetic Bloch band. It turns out that   $H^B_{\theta,q}$ is isospectral to $H^{q^2B}_{\rm Hof}$ for any $\theta$ and all spectra agree with the  Hofstadter spectrum depicted in his famous (black and white) butterfly.
However, the resulting Chern numbers of subbands, corresponding to Hall conductivities,  depend on~$\theta$ and $q$, and thus the models lead to different colored butterflies.\end{abstract}

\newpage

\section{Introduction}

We consider perturbations of the self-adjoint Schr\"odinger operator
\[
H_{B_0,\Gamma}=\tfrac12 (-\I\nabla_x - A_0 )^2 + V_\Gamma   \,,
\]
densely defined on $L^2(\R^2)$, where  $A_0:\R^2\to\R^2$ and $V_\Gamma:\R^2\to \R$ act as multiplication operators. Here  $A_0(x)=    (-B_0x_2,0)$ is the vector potential of a constant magnetic field $B_0\in\R$ and the  scalar potential $V_\Gamma$ is assumed to be periodic with respect to a Bravais lattice $\Gamma\subset \R^2$. 
The spectral properties of the operator $H_{B_0,\Gamma}$ are extremely sensitive to the relation between the numerical value of $B_0\in\R$ and the area $|\Gamma|$ of one lattice cell. 
 When $B_0$ and $\Gamma$ are commensurable in the sense that $B_0 |\Gamma|/2\pi =\frac{p}{q} \in   \Q$, the operator $H_{B_0,\Gamma}$ is unitarily equivalent by an explicit unitary transformation $\BF_q$ to a countable direct sum of multiplication operators by real-valued continuous  functions $E_n:\mathbb{T}^*_q\to \R$  with $E_n(k)\leq E_{n+1}(k)$ for all $k\in \T^*_q$ and $n\in\N=\{1,2,\ldots\}$. Here the   two-dimensional torus $\mathbb{T}^*_q$ is the Pontryagin dual of a subgroup $\Gamma_q$ of $\Gamma$. In summary it holds that
 \begin{equation}\label{directsum}
\widehat H_{B_0,\Gamma} := \BF_q \,H_{B_0,\Gamma}\, \BF^*_q = \sum_{n=1}^\infty E_n \,P_n \qquad\mbox{on }\quad \Hi := \BF_q L^2(\R^3)= L^2(\T^*_q; \Hi_{\rm f})\cong\bigoplus_{n=1}^\infty L^2(\mathbb{T}^*_q)\,,
 \end{equation}
 where $P_n$ is the orthogonal projection onto the $n$th summand in the direct sum.
As a consequence,  the spectrum $\sigma(H_{B_0,\Gamma})= \bigcup_{n=1}^\infty E_n(\T_q^*)$ is a union of intervals and purely absolutely continuous.
If, on the other hand, $B_0 |\Gamma|/2\pi \notin  \Q$, then it is expected that $\sigma(H_{B_0,\Gamma})$ is a set of Cantor-type, i.e.\ a closed nowhere-dense set of zero  Lebesgue measure. The proof of this so-called Ten-Martini problem was given only recently \cite{AJ09} and it only applies to   simple tight-binding models on $\ell^2(\Z^2)$. The most prominent picture of this commensurability problem is the fractal Hofstadter butterfly, a plot of the spectrum of such a simple tight binding model as a function of the magnetic field $B_0$, cf.\ Figure~\ref{BW} in Section~7.

The physical meaning of the  operator $H_{B_0,\Gamma}$ is that of a  Hamiltonian  for a single particle constrained to move   in a 
planar two-dimensional crystalline lattice under the influence of a constant magnetic field of strength  $  B_0$ perpendicular to the plane.  However, from the point of view of physical applications and experiments, a constant magnetic field  $B_0$ is  a highly idealized situation that can be realized only approximately. The distinction between rational and irrational magnetic fields $B_0$ is a purely mathematical one.
Thus  it is of genuine interest to understand perturbations of $H_{B_0,\Gamma}$ by potentials 
$A^\epsi(x) := A(\epsi x)$ and $\Phi^\epsi(x) := \Phi(\epsi x)$ corresponding to magnetic and electric fields 
$B^\epsi(x) := \epsi ({\rm curl} A)(\epsi x)$ and $\mathcal{E}^\epsi(x):= \epsi (\nabla \Phi)(\epsi x)$ 
that are small and slowly varying in the asymptotic limit $\epsi\ll 1$.  Here $A:\R^2\to \R^2 $ and $\Phi:\R^2\to\R $ are smooth functions.
We therefore consider the self-adjoint Schr\"odinger operator 
\[
H^\epsi_{B_0,\Gamma}=\tfrac{1}{2}(-\I\nabla_{x}-A_0 - A^\epsi)^{2}+V_{ \Gamma} +\Phi^\epsi
\]
for a fixed rational value of $B_0|\Gamma|/2\pi = \frac{p}{q}$ in the asymptotic limit $\epsi\ll 1$ as a perturbation of 
the simple block-structure \eqref{directsum}. It follows by well known techniques of adiabatic perturbation theory  that  parts of  the block-decomposition  \eqref{directsum} are stable  under such perturbations: Assuming  e.g.\ for a single function $E_n$ the gap condition $E_{n-1}(k)<E_n(k)<E_{n+1}(k)$ for all $k\in \T^*_q$, one can construct from $P_n$  an orthogonal projection $\Pi^\epsi_n$ such that $\|[\Pi_n^\epsi, \widehat H^\epsi_{B_0,\Gamma}]\|_{\mathcal{L}(\Hi)} = \Or(\epsi^\infty)$. While  the restriction $P_n\widehat H_{B_0,\Gamma}P_n$ of the  unperturbed operator  to one of its invariant subspaces  ran$P_n$ acts as multiplication   by the function $E_n$, the 
restriction $\Pi^\epsi_n\,\widehat H^\epsi_{B_0,\Gamma}\,\Pi^\epsi_n$ of the perturbed operator $\widehat H^\epsi_{B_0,\Gamma}$ to one of its almost invariant subspaces ran$\Pi^\epsi_n$ has a priori no simple form. The ``Peierls substitution rule'', widely used in physics, suggests that $\Pi^\epsi_n\,\widehat H^\epsi_{B_0,\Gamma}\,\Pi^\epsi_n$ is unitarily equivalent to a pseudodifferential operator with principal part
\[
E_n( k - A(\I\epsi\nabla_k)) + \Phi(\I\epsi\nabla_k)
\]
acting on some space of functions on the torus $\T^*_q$.  The main result of our paper is to turn this claim into a precise statement and to prove it: We show that   the blocks $\Pi^\epsi_n\,\widehat H^\epsi_{B_0,\Gamma}\,\Pi^\epsi_n$ of the perturbed operator are unitarily equivalent to pseudodifferential operators acting on   spaces of sections of possibly nontrivial vector bundles over the torus with principal part given by the  Peierls substitution rule.
A special case of our main result Theorem~\ref{5.11} is  the following statement.
\begin{thm}\label{THMintro} Let $A,\Phi$ be smooth bounded functions with bounded derivatives of any order and  $B_0 |\Gamma|/2\pi =\frac{p}{q} \in   \Q$. For any simple Bloch function $E_n$ of the unperturbed Hamiltonian $H_{B_0,\Gamma}$ satisfying the gap condition 
 there exist for  $\epsi>0$ small enough
 \begin{itemize}
 \item an orthogonal projection $\Pi^\epsi_n$,
 \item  a line bundle $\Xi_\theta$ over the torus $\T^*_q$ with connection $\nabla^\theta$ and Chern number $\theta\in\Z$, 
 \item  a unitary map $U^\epsi: {\rm ran}\Pi^\epsi_n\to L^2(\Xi_\theta)$,
 \item and a pseudodifferential operator $E^\epsi_n \in \mathcal{L}(L^2(\Xi_\theta))$ with 
 \[
 \left\| E^\epsi_n  -  \Big( E_n\left( k - A(\I\epsi\nabla^\theta_k)\right) + \Phi(\I\epsi\nabla^\theta_k)\Big)\right\|_{\mathcal{L}(L^2(\Xi_\theta))}\;=\;  \Or(\epsi)
 \]
 \end{itemize} such that $\|[\Pi_n^\epsi, \widehat H^\epsi_{B_0,\Gamma}]\|_{\mathcal{L}(\Hi)} = \Or(\epsi^\infty)$ and 
 \begin{equation}\label{PS}
\left\| U^{\epsi } \Pi^\epsi_n \,\widehat H^\epsi_{B_0,\Gamma}\,\Pi^\epsi_n  U^{\epsi *} -  E^\epsi_n \right\|_{\mathcal{L}(L^2(\Xi_\theta))}\;=\;  \Or(\epsi^\infty)\,.
\end{equation}
\end{thm}
In Theorem~\ref{5.11} we actually consider a more general situation, where a single band $E_n$ is replaced by a finite family of bands. Then  $\Xi_\theta$ becomes a vector-bundle of finite rank and the Peierls substitution Hamiltonian   is a pseudodifferential operator with matrix-valued symbol. We also compute the subprincipal symbol of $E^\epsi_n$ explicitly, which contains important information for transport and  magnetic properties of electron gases in periodic media.

Theorem~\ref{5.11}, and its special case Theorem~\ref{THMintro}, were shown before for the case $B_0=0$ in \cite{PST03b}. There one has $\theta=0$ and $\Xi_0$ is a trivial vector-bundle over the torus $\T^*_q$. 
For the case $B_0\not=0$ 
the validity and the meaning of Peierls substitution, even on a purely heuristic level, were  a matter of debate (see e.g.\  \cite{Zak86,Zak91}) and, to our knowledge, not even a precise conjecture was stated in the literature.
 
Before giving more details, let us mention that the systematic or even rigorous analysis of two-dimensional systems with periodic potential and magnetic field  is a continuing theme in theoretical physics, e.g.\
\cite{Pei33,Bl62,Zak68,Hof76,TKNN82,SN99,GA03a}, and also in mathematical physics and mathematics, e.g.\  \cite{DN80a,DN80b,Nov81,Bus87,Bel88,GRT88,HS89,RB90,HKS90,HS90a,HS90b,Nen91,GMS91,HST01,PST03b,DGR04,P07,AJ09,DP10,DL11,StT13}. We can mention here only a small part of the enormous  literature and for a  review of the mathematical and physical literature until 1991 we refer to \cite{Nen91}.
  
Most of the mathematical literature  is concerned with 
the problem of recovering the spectrum and sometimes the density of states of the perturbed  Hamiltonian $H^\epsi_{B_0,\Gamma}$. In some cases this is done by constructing isospectral effective Hamiltonians in spirit of the Peierls substitution rule, see e.g.\ \cite{HS89,RB90,HKS90,HS90a,HS90b,GMS91}. With a few exceptions, most notably \cite{RB90}, 
the limiting cases $B_0=0$ and $B_0\to \infty$  were considered.
More recently the question of constructing unitarily equivalent effective Hamiltonians was taken up in \cite{PST03b,DP10,DL11} and the limiting regimes $B_0=0$ and $B_0\to \infty$ are fully understood by now even on a mathematical level. For a thorough discussion   of the question why unitary equivalence is important also from a physics point of view,  we refer to  \cite{DP10}. Let us mention here only one example: The two canonical models for effective Hamiltonians  for the asymptotic regimes  $B_0=0$ and $B_0\to \infty$ are exactly isospectral. This is known as the duality of the Hofstadter model, see e.g.\ \cite{GA03b}. However, they are not unitarily equivalent and describe different physics.

The problem of constructing unitarily equivalent effective Hamiltonians in the intermediate regime of finite $B_0\not=0$ was, to our knowledge,  completely open up to now\footnote{
It was observed in \cite{DGR04} that the method of \cite{PST03b} can be directly applied also to magnetic Bloch bands if one assumes that the magnetic Bloch bundles are trivial. But this assumption is generically not satisfied.} and its solution is the main content of our paper.
 While we  use the same basic approach that was applied in \cite{PST03b,DP10} for the cases $B_0=0$ and $B_0\to \infty$, namely adiabatic perturbation theory \cite{PST03a}, 
there is a major geometric obstruction in extending these methods to perturbations around finite   values of $B_0$ such that $B_0 |\Gamma|/2\pi =\frac{p}{q} \in  \Q$, which we shortly explain. 
In all cases the projections $P_n$ in \eqref{directsum} act on $L^2(\T^*_q,\Hi_{\rm f})$  fiber-wise, i.e.\ they are given by projection-valued functions $P_n:\T^*_q\to \mathcal{L}(\Hi_{\rm f})$, $k\mapsto P_n(k)$. For an isolated simple  band $E_n$ the corresponding projection-valued function $P_n(\cdot)$ is   smooth and   defines a complex line-bundle over $\T^*_q$, the so called Bloch bundle associated with the Bloch band $E_n$. 
For $B_0=0$ the Bloch bundles  are trivial 
and the effective operator $E_n^\epsi$ is a pseudodifferential operator acting on $L^2(\mathbb{T}^*_1)$, the space of $L^2$-sections of the trivial line bundle over the torus $\mathbb{T}^*_1$. 
The 
Bloch bundles for $B_0\not=0$ are not trivial in general and  $E_n^\epsi$ has to be understood as a pseudodifferential operator acting on the sections of a non-trivial line bundle $\Xi_\theta$ over the torus $\mathbb{T}^*_q$.

An important shortcoming of our result is, however, that  we can not allow for the case of a perturbation by a constant magnetic field $B$, corresponding to  a linear vector potential~$A$, in all steps of the derivation. While an (almost) invariant subspace and the corresponding (almost) block structure of the perturbed Hamiltonian can still be established   in this case, and also the effective Hamiltonian  ${\rm Op}^\theta (E_n(k -   A(r))+ \Phi(r))$ remains well defined for linear~$A$, the unitary map  intertwining the (almost) invariant subspace and the reference space, as we construct it, no longer exists. For $\theta=0$ this problem actually disappears, and we recover the results for non-magnetic Bloch bands with constant small magnetic fields $B$ obtained in \cite{DP10,DL11}. 
Note, however, that the physically relevant situation where $B$ and also $E=- \nabla\Phi$ are constant over a macroscopic volume containing $\epsi^{-2}$ lattice sites  is included in all of our results.

Let us mention that some of the physically relevant questions can be answered without establishing Peierls substitution in our sense of unitary equivalence. There are, in particular, semiclassical and algebraic approaches that allow for direct computation of many relevant quantities without the detour via Peierls substitution. The modified semiclassical equations of motion  for magnetic Bloch bands  \cite{SN99} became the starting point for a large number of quantitative results, see e.g.\ \cite{XCN10} and references therein.
This approach was rigorously derived and extended in \cite{StT13,Teu12}. 
In \cite{GA03a} the authors apply Bohr-Sommerfeld quantization with phases modified by the Berry curvature and the Rammal-Wilkinson term in order to 
compute the orbital magnetization in the Hofstadter model. For the case $B=$const.\ or periodic and $\Phi=0$ the algebraic approach of Bellissard and coworkers  \cite{Bel88,RB90,BKS91} provides a powerful tool  for expansions to all orders for eigenvalues, free energies and quantities  derived from there. This approach can also cope with random perturbations and has developed into a very general machinery, see e.g.\  \cite{BES94, ST13} and references therein.

We end the introduction with a short outline of the paper. In Section~2 we give a precise formulation of the setup and introduce all relevant quantities and assumptions. In Section~3 we briefly formulate the result on the existence and the construction of almost invariant subspaces. We do not give a proof here, since nothing interesting changes with respect to the non-magnetic case at this point. In Section~4 we analyze in detail the structure of magnetic Bloch bundles. As a result we can construct the reference space for the effective Hamiltonian and the unitary map  from the almost invariant subspace to this reference space. This analysis is one key   ingredient to our main result, which we formulate and prove in Section~5. The result and its proof are based  on   geometric Weyl calculi for operators acting on sections of non-trivial vector bundles, the other  key ingredients, that are developed in Section~6. 
In the final Section~7, we explicitly compute Peierls substitution Hamiltonians for magnetic subbands of the Hofstadter Hamiltonian. The Hofstadter model is the canonical model for a single non-magnetic Bloch band perturbed by a constant magnetic field $B_0$. As a result we find a new two-parameter family $H^B_{\theta,q}$ (see \eqref{Hthetaq}) of Hofstadter like Hamiltonians
indexed by integers $\theta\in \Z$ and $q\in \N$. The operator $H^B_{\theta,q}$ can be viewed as the canonical model for a magnetic Bloch band with Chern number $\theta$ and originating from a Bloch band split into $q$ magnetic subbands. Like the Hofstadter model itself, all  $H^B_{\theta,q}$ are representations of an element of the non-commutative torus algebra, the abstract Hofstadter operator. As a consequence they are all isospectral and lead to the same black and white butterfly, Figure~\ref{BW}. But the transport properties encoded in the Chern numbers of spectral bands depend on $\theta$ and $q$ and they give rise to different colored butterflies, cf.\ Figure~\ref{subbut}. The results of Section~7 and a more detailed analysis presented in \cite{ADT15} suggest that our main Theorem~\ref{5.11}   also holds for perturbations by   magnetic fields with potentials $A$ of linear growth. 
 \medskip

\noindent {\bf Acknowledgment:} We are 
grateful  to Abderram\'an Amr for his involvement in a related project \cite{Amr15,ADT15} which had important impact on   Section~7. In particular, his code was used to produce Figure~\ref{subbut}.
We thank Giuseppe De Nittis, 
  Jonas Lampart, Gianluca Panati,  and Jakob Wachsmuth for numerous very helpful discussions and
 for continued exchange about many questions closely related to the content of this work.
We thank Max Lein  for  his careful  reading of a preliminary version of the manuscript.

\section{Perturbed periodic and magnetic Schr\"odinger operators}

We consider perturbations of a one-particle Schr\"odinger operator with a periodic potential and a    constant magnetic field   in two dimensions.  
The unperturbed operator is given by
\begin{equation*}\label{H_MB}
H_{\mathrm{MB}}=\tfrac{1}{2}(-\I\nabla_{x}-A^{(0)}(x))^{2}+V_{\tilde \Gamma}(x)
\end{equation*}
with domain $H^2_{A^{(0)}}(\R^{2})$, a magnetic Sobolev space. Here 
\[
A^{(0)}(x):=\ \mathcal{B}_0 x\quad\mbox{ with } \quad \mathcal{B}_0 := \begin{pmatrix} 0 & -B_0\\ 0&0\end{pmatrix}
\]
 and  $V_{\tilde \Gamma}$ is periodic with respect to a Bravais lattice 
\[
\tilde \Gamma := \left\{ a\tilde  \gamma_1 + b\tilde \gamma_2 \in \R^2\,|\, a,b \in\Z 
\right\}
\]
spanned by a basis $(\tilde \gamma_1,\tilde \gamma_2)$ of $\R^2$,
 i.e. $V_{\tilde \Gamma}(x+\tilde \gamma)=V_{\tilde \Gamma}(x)$
for all $\tilde \gamma \in \tilde \Gamma$. 
We will later assume that $B_0\in\R$ satisfies a commensurability condition, so that $H_{\rm MB}$ obtains a magnetic Bloch band structure.

The full Hamiltonian is a perturbation of $H_{\rm MB}$ by ``small'' magnetic and electric fields of order $\epsi$. More precisely let $A^{(1)} $ be a linear vector potential of an additional constant magnetic field $B_1$ and $A^{(2)}$ and $\Phi$ be bounded vector and scalar potentials, then   the full Hamiltonian $H^{\varepsilon}$   reads
\begin{equation}\label{H_eps}
H^{\varepsilon}=\tfrac{1}{2}(-\I\nabla_{x}-A^{(0)}(x)- \epsi  A^{(1)}(  x)-A^{(2)}(\varepsilon x))^{2}+V_{\tilde \Gamma}(x)+\Phi(\varepsilon x)
\end{equation}
  with domain $H^2_{A^{(0)}+\epsi A^{(1)}}(\R^{2})$, where 
  \[
  H^m_A := \{ f\in L^2(\R^2)\,|\, (\I\nabla_x + A(x))^\alpha f\in L^2(\R^2) \mbox{ for all }  \alpha\in \N_0^2\mbox{ with }|\alpha| \leq m   \} 
  \]
  and $\N_0= \{0,1,2,\ldots\}$.

\begin{ass}\label{A_1: A, Phi, V_Gamma}
Assume that $A^{(2)} \in C^{\infty}_{\mathrm{b}}(\R^{2},\R^{2})$ satisfies the gauge condition $A^{(2)}(x)\cdot \tilde\gamma_2 = 0$   for all $x\in\R^2$ and that $\Phi \in C^{\infty}_{\mathrm{b}}(\R^{2},\R)$. Let   $V_{\tilde \Gamma}:\R^2\to\R$ be a measurable function such that $V_{\tilde \Gamma}(x+\tilde \gamma)=V_{\tilde \Gamma}(x)$
for all $\tilde \gamma \in \tilde \Gamma$ and that the operator of multiplication by $V_{\tilde \Gamma}$  is relatively $(-\I\nabla-A^{(0)}-\epsi A^{(1)})^{2}$-bounded with relative bound smaller than $1$ for all $\epsi>0$ small enough. 
\end{ass}

Under these conditions, $H_{\ro{MB}}$ and $H^{\eps}$ are essentially self-adjoint   
on $C_0^\infty(\R^2)$ and self-adjoint on 
  $H^2_{A^{(0)} }(\R^{2})$ respectively on $H^2_{A^{(0)}+\epsi A^{(1)}}(\R^{2})$.
Note that any $V_{\tilde \Gamma}\in L^2_{\rm loc}(\R^2)$ satisfies Assumption~\ref{A_1: A, Phi, V_Gamma}.

\subsection{The  band structure of $H_{\rm MB}$}

The magnetic translation  of functions on $\R^2$ by $\tilde \gamma_j$ is defined by
\begin{equation}\label{magn translations}
(\tilde T_j \psi)(x):=  \E^{\I\langle x,\mathcal{B}_0\tilde \gamma_j \rangle }\psi(x-\tilde \gamma_j)\,.
\end{equation}
On $L^2(\R^2)$ the magnetic translations are unitary and leave invariant 
  the magnetic momentum operator and the periodic potential\,,
\[
\tilde T_j^{-1} \,(-\I\nabla - A^{(0)})\,\tilde T_j = (-\I\nabla - A^{(0)})\,, \quad \tilde T_j^{-1} \,V_{\tilde \Gamma}\,\tilde T_j = V_{\tilde \Gamma}\quad\mbox{and thus}\quad \tilde T_j^{-1} \,H_{\rm MB}\,\tilde T_j = H_{\rm MB}.
\]
Because of
\[
\tilde T_1\tilde T_2 =\E^{ \I\langle \tilde \gamma_{2},\mathcal{B}_0 \tilde \gamma_{1}\rangle } \tilde T_2\tilde T_1\,,
\]
 we only obtain a unitary representation of $\tilde \Gamma$ if $\langle \tilde \gamma_{2},\mathcal{B}_0 \tilde \gamma_{1}\rangle\in 2\pi \Z$. Here $\langle \tilde \gamma_{2},\mathcal{B}_0 \tilde \gamma_{1}\rangle = B_0  \tilde \gamma_1 \wedge \tilde \gamma_2$ is   the magnetic flux through the unit cell $M$ of the lattice $\Gamma$ with oriented  volume $\tilde \gamma_1 \wedge \tilde \gamma_2$.

 \begin{ass}\label{A_2: B0rational}
 The flux of $B_0$ per unit cell satisfies $\langle \tilde \gamma_{2},\mathcal{B}_0 \tilde \gamma_{1}\rangle = 2\pi\frac{p}{q} \in  2\pi \mathbb{Q}$.
 \end{ass}

By passing to the      sublattice $  \Gamma\subset\tilde \Gamma$  spanned by the basis  $(\gamma_1,\gamma_2):=(q\tilde \gamma_1,\tilde \gamma_2)$ and defining the magnetic translations $T_1$, $T_2$ analogously, we achieve 
 $\langle   \gamma_{2},\mathcal{B}_0   \gamma_{1}\rangle= 2\pi p \in 2\pi \Z$. Hence
 \begin{equation}\label{TDef}
 T: \Gamma \to \mathcal{L}(L^2(\R^2))\,,\quad \gamma = n_1\gamma_1+n_2\gamma_2 \;\;\mapsto \;\;
T_\gamma :=  T_1^{n_1}T_2^{n_2}
 \end{equation}
is a unitary representation of $\Gamma$ on $L^2(\R^2)$ satisfying 
\begin{equation}\label{THcommu}
T_\gamma^{-1} H_{\rm MB}T_\gamma = H_{\rm MB}
\end{equation}
 for all $\gamma\in\Gamma$.
Before we introduce the Bloch-Floquet transformation in order to exploit the translation invariance of $H_{\rm MB}$,   we first define  a number of useful function spaces. Let
$$\mathcal{H}_{\mathrm{f}}:=\left\{ f \in L^{2}_{\mathrm{loc}}(\R^{2})\,|\,T_{\gamma} f=f\quad\mbox{for all}\quad \gamma \in \Gamma\right\}\,,$$
which, equipped with the inner product
$\langle f,g \rangle _{\mathcal{H}_{\mathrm{f}}}:=\int_{M}\overline{f(y)}g(y)\mathrm{d}y$,
  is a Hilbert space. Analogously for   $m \in \N$ 
$$\mathcal{H}^{m}_{A^{(0)}}(\R^{2}):=\left\{f\in \mathcal{H}_{\mathrm{f}}\,|\,(-\I\nabla-A^{(0)})^{\alpha}f \in \mathcal{H}_{\mathrm{f}} \quad\mbox{for all $\alpha\in \N_0^2$ with $|\alpha| \leq m$} \right\}$$
is a Hilbert space with inner product
\[
\langle f,g\rangle_{\mathcal{H}^{m}_{A^{(0)}}(\R^{2})}:=\sum_{|\alpha|\leq m} \langle (-\I\nabla-A^{(0)})^{\alpha}f,(-\I\nabla-A^{(0)})^{\alpha}g\rangle_{\mathcal{H}_{\mathrm{f}}}\,.
\]

Let $\Gamma^{*} $ be the dual lattice of $\Gamma $, i.e.\ the $\Z$-span of   the unique basis $(\gamma_1^*,\gamma_2^*)$ such that $\gamma_i^*\cdot \gamma_j = 2\pi \delta_{ij}$. By    $M$ respectively $M^{*}$ we denote the centered fundamental cells of $\Gamma$ respectively of~$\Gamma^*$.
On $\mathcal{H}_{\rm f}$ a unitary representation of the dual lattice  $\Gamma^*$ is given by
\[
\tau: \Gamma^*\to \mathcal{L}(\mathcal{H}_{\rm f})\,,\quad \gamma^* \mapsto \tau(\gamma^{*})
\quad\mbox{ with } (\tau(\gamma^{*})f)(y):=\E^{\I y\cdot \gamma^{*}}f(y)\,.
\]
Finally let the space of $\tau$-equivariant functions be 
$$\mathcal{H}_{\tau}:=\{ f \in L^{2}_{\mathrm{loc}}(\R^{2}_{k}, \mathcal{H}_{\mathrm{f}} )\,|\,f(k-\gamma^{*})=\tau(\gamma^{*})f(k) \quad\text{for all}\quad \gamma^{*} \in \Gamma^{*}\}\,.
$$
Equipped with the inner product
$\langle f,g\rangle_{\mathcal{H}_{\tau}}=\int_{M^{*}}\langle f(k),g(k)\rangle_{\mathcal{H}_{\mathrm{f}}} \mathrm{d}k,$
where $\mathrm{d}k$ is the normalized Lebesgue measure on $M^*$, $\mathcal{H}_\tau$ is a Hilbert space.

For $\psi\in C^\infty_0(\R^2)$ the magnetic  Bloch-Floquet transformation is defined by 
\begin{equation}\label{UBF}
(\mathcal{U}_{\mathrm{BF}}\psi)(k,y) := \sum_{\gamma \in \Gamma} \E^{-\I(y-\gamma)\cdot k}(T_{\gamma}\psi)(y)\,.
\end{equation}
It extends uniquely to a unitary mapping 
$\mathcal{U}_{\mathrm{BF}}: L^{2}(\R^{2})\rightarrow \mathcal{H}_{\tau}$
 and its inverse   is given by
$$(\mathcal{U}_{\mathrm{BF}}^{-1}\phi)(x)=\int_{M^{*}}\E^{\I k\cdot x} \phi(k,x) \mathrm{d}k.$$
Because of \eqref{THcommu} the operator  $H_{\ro{MB}}$ fibers in the magnetic Bloch-Floquet representation as
$$H^{0}_{\mathrm{BF}}:=\mathcal{U}_{\mathrm{BF}}\, H_{\mathrm{MB}}\, \mathcal{U}_{\mathrm{BF}}^{*} = \int_{M^{*}}^{\oplus}H_{\mathrm{per}}(k) \,\mathrm{d}k\,,$$
where
\begin{equation*}\label{H_per}
H_{\mathrm{per}}(k):=\tfrac{1}{2}(-\I\nabla_{y}-A^{(0)}(y)+k)^{2}+V_{\Gamma}(y)
\end{equation*}
acts for any fixed $k\in M^*$ on the  $k$-independent domain $\mathcal{H}^{2}_{A^{(0)}}(\R^{2}) \subset \mathcal{H}_{\rm f}$.
The domain $H^2_{A^{(0)}}(\R^2)$ of $H_{\rm MB}$ is mapped to 
\[
\mathcal{U}_{\rm BF} H^2_{A^{(0)}}(\R^2)=:L^{2}_{\tau}(\R^{2},\mathcal{H}^{2}_{A^{(0)}}(\R^{2}))= L^{2}_{\mathrm{loc}}(\R^{2},\mathcal{H}^{2}_{A^{(0)}}(\R^{2}))\cap \mathcal{H}_\tau\,.
\]
  As $H_{\mathrm{per}}(k)$ basically describes a Schr\"odinger particle in a box, it is   bounded from below and 
  has a compact resolvent for every $k \in M^{*}$. Hence $H_{\mathrm{per}}(k)$ has discrete spectrum with eigenvalues $E_n(k)$ of finite multiplicity that accumulate at infinity.
So let
$$E_{1}(k) \leq E_{2}(k) \leq...$$
be the eigenvalues   repeated according to their multiplicity. In the following, $k\mapsto E_{n}(k)$ will be called the $n^{\ro{th}}$ band function or just the $n^{\ro{th}}$ Bloch band.
Since $H_{\rm per}(k)$ is $\tau$-equivariant, i.e.\
\[
H_{\rm per}(k-\gamma^*) = \tau(\gamma^*) \,H_{\rm per}(k)\,\tau(\gamma^*)^{-1}\,,
\]
and $\tau(\gamma^*)$ is unitary, the Bloch bands $E_n(k)$ are $\Gamma^*$-periodic functions.

\begin{figure}\begin{center}
\includegraphics[height=5cm]{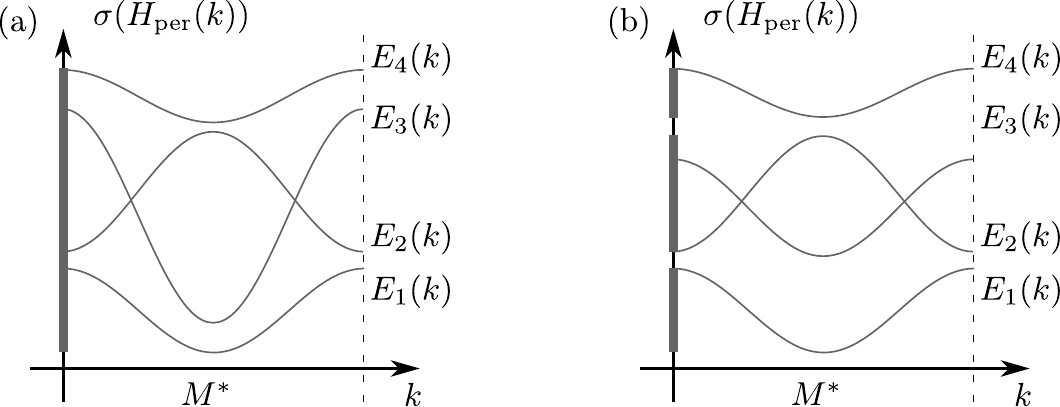}\end{center}\vspace{-3mm}

\caption{\small Two sketches of Bloch bands. Note that   $k\in\R^2$, so the graphs of the Bloch bands are really surfaces. In (a) the families  $\{E_1(k)\}$,  $\{E_2(k),E_3(k)\}$ and $\{E_4(k)\}$ are all isolated, but none of them is strictly isolated.  In (b) they are all strictly isolated.}
\label{blochbandsfig}
\end{figure}

 The effective Hamiltonians that we construct will be associated with isolated families of Bloch bands of the unperturbed operator $H_{\rm per}(k)$.
\begin{defi}\label{gap}
A family of bands $\{E_{n}(k)\}_{n\in I}$ with $I=[I_{-},I_{+}]\cap \N$ is called isolated, or synonymously is said to satisfy the gap condition, if
$$\inf_{k\in M^{*}} \ro{dist} \left(\cup _{n\in I}\{E_{n}(k)\},\cup_{m \notin I}\{E_{m}(k)\}\right)=:c_{\mathrm{g}} >0.$$
We say that $\{E_{n}(k)\}_{n\in I}$ is strictly isolated with strict gap $d_{\rm g}$ if for 
\[
\sigma_I := \overline{\cup_{n\in I }\cup_{k\in M^{*}}\{E_n(k)\}}\
\]
we have that 
$$\inf_{m\notin I, k\in M^*} \mathrm{dist}(  E_m(k), \sigma_I ):=d_{\ro{g}}>0\,.$$
\end{defi}

By $P_I(k)$ we denote the spectral projection of $H_{\rm per}(k)$ corresponding to the isolated family of eigenvalues $\{E_{n}(k)\}_{n\in I}$. Because of the gap condition, the map 
\[
\R^2\to \mathcal{L}(\mathcal{H}_{\rm f})\,,\quad k\mapsto P_I(k)
\]
is real analytic and 
with $H_{\rm per}(k)$ also $\tau$-equivariant.
This family of projections defines a vector bundle over the torus $\mathbb{T}^* := \R^2/ \Gamma^*$.
\begin{defi}
Let the   bundle $\pi:\bu_\tau\to \mathbb{T}^*$ with typical fiber $\Hi_{\rm f}$ be given by
\[
\bu_\tau:= ( \R^{2}\times \mathcal{H}_{\mathrm{f}} )/_{\displaystyle \sim_\tau},
\]
where
$$  (k,\varphi) \sim_\tau (k',\varphi')\quad :\Leftrightarrow \quad k'=k-\gamma^{*} \quad\mbox{and}\quad \varphi '=\tau(\gamma^{*})\varphi\quad\mbox{for some $\gamma^*\in\Gamma^*$}\,.$$
The Bloch bundle $\bu_{\rm Bl}$ associated to the isolated family $\{E_{n}(k)\}_{n\in I}$ of Bloch bands is the subbundle given by 
\begin{equation}\label{E_Bl}
\bu_{\mathrm{Bl}}:=\{(k,\varphi) \in \R^{2}\times \mathcal{H}_{\mathrm{f}}  \,|\, \varphi \in P(k)\mathcal{H}_{\mathrm{f}}\}/_{\displaystyle \sim_\tau}\,.
\end{equation}

\end{defi}
Hence, the $L^2$-sections of $\bu_\tau$ are in one-to-one correspondence with elements of $\Hi_\tau$ and the $L^2$-sections of
the Bloch bundle are in one-to-one correspondence with  functions $f\in\mathcal{H}_\tau$ that satisfy $P_I(k)f(k)=f(k)$ for all $k\in\R^2$.

\subsection{$H^\epsi$ as a pseudodifferential operator on $\mathcal{H}_\tau$}

The operator of multiplication by $x$ on $L^2(\R^2)$ is mapped under the Bloch-Floquet transformation to the operator $\I \nabla_k^\tau := \mathcal{U}_{\rm BF} \,x\,\mathcal{U}_{\rm BF}^*$. A simple computation shows that $\I \nabla_k^\tau $ acts  as the gradient  with domain 
$H^1_{\rm loc} (\R^2, \mathcal{H}_{\rm f}) \cap \mathcal{H}_\tau\subset  \mathcal{H}_\tau$.
Hence, by the functional calculus for self-adjoint operators, the full Hamiltonian $H^\epsi$ takes the form 
\begin{equation*}\label{H_epsBF}
H^{\eps}_{\mathrm{BF}}:=\mathcal{U}_{\mathrm{BF}} H^{\eps} \mathcal{U}_{\mathrm{BF}}^{*}= \tfrac{1}{2}(-\I\nabla_{y}-A^{(0)}(y)+k-  A (\I\eps \nabla_{k}^{\tau}) )^{2}+V_{\Gamma}(y)+\Phi(\I\eps \nabla_{k}^{\tau})\,,
\end{equation*}
where   we put $A:= A^{(1)}+A^{(2)}$ and use that  $\epsi A^{(1)}(x) = A^{(1)}(\epsi x) $ due to linearity.
One key step for the following analysis is to interpret $H^{\eps}_{\mathrm{BF}}$ as a pseudodifferential operator with operator valued symbol
\begin{equation}\label{symbol H_eps}
H(k,r):=\tfrac{1}{2}(-\I\nabla_{y}-A^{(0)}(y)+k-A(r))^{2}+V_{\Gamma}(y)+\Phi(r)
\end{equation}
under the quantization map $k\mapsto k$ and $r\mapsto \I\epsi\nabla_k^\tau$.
To make this precise, note that $H(k,r)$ is a $\tau$-equivariant symbol taking values in the 
self-adjoint operators on $\mathcal{H}_{\rm f}$ with domain $\mathcal{H}^{2}_{A^{(0)}}$ independent of $(k,r)$. For convenience of the reader we briefly give the definitions of the relevant symbol classes and refer to the Appendix~B of \cite{Te03}   for details on the $\tau$-quantization.

\begin{defi}
A  function $w:\R^{4}\to [0,\infty)$ satisfying, for some $C,N>0$, 
\[
w(x)\leq C \langle x-y\rangle^N w(y)\,,\quad\forall x,y\in\R^4\,,
\]
is called order function. Here $\langle x\rangle := (1+|x|^2)^\frac12$.\\[2mm]
Let $\Hi_1$ and $\Hi_2$ be Hilbert spaces and $w$ an order function. Then by $S^w(\mathcal{L}(\Hi_1,\Hi_2))$ we denote  the  space  functions $f\in C^\infty(\R^{4}, \mathcal{L}(\Hi_1,\Hi_2))$,  that satisfy 
\[
\|f\|_{w,\alpha,\beta}:= \sup_{(k,r)\in\R^{4}}  {w(k,r)^{-1}} \| (\partial_k^\alpha \partial_r^\beta f)(k,r)\|_{\mathcal{L}(\Hi_1,\Hi_2)}  < \infty   \quad\mbox{ for all $\alpha,\beta \in \N^2_0$.}
\]
Functions in $S^w(\mathcal{L}(\Hi_1,\Hi_2))$ are called  operator-valued symbols with order function $w$.
 For  the constant order function $w(k,r)\equiv 1$ we write  $S^{1} (\mathcal{L}(\Hi_1,\Hi_2)):=  S^{w\equiv 1}(\mathcal{L}(\Hi_1,\Hi_2))$. \\[2mm]
Let $\tau_j: \Gamma^*\to \mathcal{L}(\Hi_j)$, $j=1,2$, be unitary representations.
A symbol $f\in S^w(\mathcal{L}(\Hi_1,\Hi_2))$ is called $(\tau_1,\tau_2)$-equivariant, 
if
\[
f(k-\gamma^*,r) = \tau_2(\gamma^*) \,f(k,r)\,\tau_1(\gamma^*)^{-1} \quad \mbox{for all $\gamma^*\in\Gamma^*$ and $(k,r)\in \R^4$.}
\]
The corresponding space is denoted by $S^w_{(\tau_{1},\tau_{2})}(\mathcal{L}(\Hi_1,\Hi_2))$ and equipped with the Fr\'echet metric induced by the family of  semi-norms $\|\cdot\|_{w,\alpha,\beta}$. \\[2mm]The space of uniformly bounded functions $f:[0,\epsi_0)\to S^w_{(\tau_{1},\tau_{2})}(\mathcal{L}(\Hi_1,\Hi_2))$ is denoted by $S^w_{(\tau_{1},\tau_{2})}(\epsi,\mathcal{L}(\Hi_1,\Hi_2))$. If $\Hi_1=\Hi_2=\Hi$ and $\tau_1=\tau_2$, we write $S^w_{\tau}(\epsi,\mathcal{L}(\Hi))$ instead.
\end{defi}
 
 \begin{prop} Let    $w_A(k,r) := 1 + |k-A(r)|^2$. Then 
 the operator valued  function $(k,r)\mapsto H(k,r)$ defined in \eqref{symbol H_eps} is a symbol
$H \in S^{w_A}_{(\tau_{1},\tau_{2})}( \mathcal{L}(\mathcal{H}^{2}_{A^{(0)}},\mathcal{H}_{\mathrm{f}}))$ with $\tau_{1}=\tau|_{\mathcal{H}^{2}_{A^{(0)}}}$ and $\tau_{2}=\tau$.\end{prop}
\begin{pro}
Since $H(k,r) = H_{\rm per}(k-A(r)) + \Phi(r)$, all claims can be checked explicitly on   $H_{\rm per}$ 
using Assumption~\ref{A_1: A, Phi, V_Gamma}: 
 The $(\tau_1,\tau_2)$-equivariance of $H$ follows   from the $(\tau_1,\tau_2)$-equivariance of $H_{\rm per}$ and 
 $H_{\rm per} \in S^{w_0}_{(\tau_{1},\tau_{2})}( \mathcal{L}(\mathcal{H}^{2}_{A^{(0)}},\mathcal{H}_{\mathrm{f}}))$ with $w_0 (k,r) := 1+|k|^2$ implies $H \in S^{w_A}_{(\tau_{1},\tau_{2})}(\R^{4},\mathcal{L}(\mathcal{H}^{2}_{A^{(0)}},\mathcal{H}_{\mathrm{f}}))$. See Lemma~3.8 in \cite{DP10} for details on the last argument.
\end{pro}
\smallskip

Note that the Weyl quantization of  a symbol  $f$ in $S^w_{(\tau_{1},\tau_{2})}(\mathcal{L}(\Hi_1,\Hi_2))$
defines  an  operator  ${\rm Op}^{(\tau_1,\tau_2)}( f)$ that maps  $\Hi_1$-valued $\tau_1$-equivariant functions to $\Hi_2$-valued $\tau_2$-equivariant functions. 
For details on this $\tau$-quantization see  Appendix~B of \cite{Te03}. For a general introduction to pseudodifferential operators with operator valued symbols  in the same context we refer to \cite{GMS91}.

Since the $(\tau_1,\tau_2)$-quantization ${\rm Op}^{(\tau_1,\tau_2)}(H)$ of $H$  restricted 
to the space   
of smooth $\tau$-equivariant functions with values in $\mathcal{H}^{2}_{A^{(0)}}(\R^{2})$ agrees with the restriction of $H^\epsi_{\rm BF}$, and since  both operators are essentially self-adjoint on this subspace, their  closures agree and we will identify them in the following.

\section{Almost invariant subspaces}

The first step of space-adiabatic perturbation theory is the construction of the almost invariant subspace $\Pi^\eps_I\mathcal{H}_{\tau}$ associated with an isolated family of Bloch bands $\{E_{n}(k)\}_{n\in I}$. 
Here $\Pi_I^\epsi$ is an orthogonal projection almost commuting with  $H^\epsi_{\rm BF}$.
This concept goes back to \cite{Nen02} and the general construction was introduced in  \cite{NeSo04,MaSo02} based on techniques developed already in \cite{HS90a}. 
The application to  the case of non-magnetic Bloch bands including the $\tau$-equivariant Weyl calculus was worked out  in \cite{PST03b, Te03}.  Since these methods  carry over to the   case  of magnetic Bloch bands without difficulties, see also  \cite{DGR04,Sti11}, we skip the details of the proof. Note, however, that we add a new observation to the statement: Under the assumption of a strict gap and for sufficiently small perturbations the resulting projection $\Pi_I^\epsi$ actually commutes with $H^\epsi_{\rm BF}$, since it turns out to be a spectral projection.
 
\begin{thm}\label{space-adiabatic threorem}
Let Assumptions~\ref{A_1: A, Phi, V_Gamma} and \ref{A_2: B0rational} hold  and 
let $\{E_{n}(k)\}_{n\in I}$  be an isolated family of Bloch bands.
 Then there exists an orthogonal projection $\Pi_I^{\eps} \in \mathcal{L}(\mathcal{H}_{\tau})$ such that $H^{\eps}_{\mathrm{BF}}\Pi_I^{\eps}$ is a bounded operator and
$$
\left\| [H^{\eps}_{\mathrm{BF}},\Pi_I^{\eps}]\right\|=\mathcal{O}(\eps^{\infty})\,.
$$
Moreover, $\Pi_I^{\eps}$ is close to a pseudodifferential operator ${\rm Op}^{\tau}\hspace{-1pt}(\pi)$,
\begin{equation}\label{pseudo}
\left\|\Pi_I^{\eps}-{\rm Op}^{\tau}\hspace{-1pt}(\pi)\right\|=\mathcal{O}(\eps^{\infty})\,,
\end{equation}
where $\pi \in S^{1}_{\tau}(\eps,\mathcal{L}(\mathcal{H}_{\mathrm{f}})):= S^{w\equiv 1}_{\tau}(\eps,\mathcal{L}(\mathcal{H}_{\mathrm{f}}))$ with principal symbol $\pi_{0}(k,r)=P_I(k-A(r))$.\\[2mm]
If $\{E_{n}(k)\}_{n\in I}$ is strictly isolated with gap $d_{\rm g}$ and $\left\| \Phi \right\|_{\infty} < \tfrac{1}{2}d_{\ro{g}}$, then \eqref{pseudo}  holds for 
$\Pi_I^{\eps}$ being  the spectral projection of $H^{\epsi}_{\ro{BF}}$ 
associated to the  interval $[\inf E_I - d_{\rm g}/2, \sup E_I + d_{\rm g}/2] $. In particular,
$[ H^{\epsi}_{\ro{BF}},\Pi_I^\epsi]=0$ in this case.

\end{thm}
 
\begin{pro}  The construction of $\Pi_I^{\eps}$ is given in Proposition~5.16 in \cite{Te03} for general Hamiltonians with symbol $\tilde H\in S^{w }_{(\tau_{1},\tau_{2})}( \R^4,\mathcal{L}(\mathcal{D} ,\mathcal{H}_{\mathrm{f}}))$ for $w (k,r) = 1 + |k|^2$, where $\tilde H(k,r)$ is  pointwise a self-adjoint operator on $\mathcal{H}_{\mathrm{f}}$ with domain $\mathcal{D}$. In the case $ A^{(1)}=0$ it applies verbatim also to  our Hamiltonian, since then   $H \in S^{w}_{(\tau_{1},\tau_{2})}(\R^{4},\mathcal{L}(\mathcal{H}^{2}_{A^{(0)}},\mathcal{H}_{\mathrm{f}}))$.  
The  slight modification  that allows to include also a linear term $A^{(1)}\not=0$ is worked out in Theorem~3.12 (1) in 
 \cite{DP10}, where the order function $w$ is replaced by $w_A$.   Note that  their assumption (D) on the triviality of the Bloch bundle is not used in the proof of part (1) of Theorem~3.12 in \cite{DP10}.
We remark that the construction of  $\Pi^\epsi_I$ for nonzero $A^{(0)}$ and $A^{(1)}=0$ was also done in \cite{Sti11}.

The   statement for strictly isolated bands follows from inspecting,  for example, the proof of Proposition~5.16 in \cite{Te03}, from where also the following notation is borrowed. Under the   assumption of a strict gap, the Moyal resolvent $R(\zeta)$ 
can be constructed globally on $\mathcal{U}_{z_0}=\R^4$ and for $\zeta$ in a fixed positively oriented  circle $\Lambda\subset \C$ encircling $[\inf E_I - d_{\rm g}/2, \sup E_I + d_{\rm g}/2]$.
  But then (5.28) in \cite{Te03} implies ${\rm Op}^{\tau}(R(\zeta) )= ({\rm Op}^{(\tau_1,\tau_2)}(H) -\zeta)^{-1} + \Or(\epsi^\infty)$ and thus, by~(5.32), 
\[
{\rm Op}^{\tau}(\pi) = \frac{\I}{2\pi} \oint_\Lambda {\rm Op}^{\tau}(R(\zeta))\,\D\zeta = \frac{\I}{2\pi} \oint_\Lambda (H^{\epsi}_{\ro{BF}} -\zeta)^{-1} \,\D\zeta + \Or(\epsi^\infty) 
 \,.
\]
\end{pro}

\section{Magnetic Bloch bundles}

With respect to the (almost) invariant subspace $\Pi^\epsi_I\mathcal{H}_\tau$ associated to an isolated family of Bloch   bands, the Hamiltonian  thus takes the (almost) block-diagonal form
\[
H^{\epsi}_{\ro{BF}}= \Pi^\epsi_I H^{\epsi}_{\ro{BF}} \Pi^\epsi_I + (1- \Pi^\epsi_I)H^{\epsi}_{\ro{BF}}(1- \Pi^\epsi_I) +\Or(\epsi^\infty)\,,
\]
where $\Or(\epsi^\infty)$ holds in the operator norm. For strictly isolated bands $\Or(\epsi^\infty)$ can be replaced by zero and the addendum ``almost'' can be dropped. The remaining task is to show that the block $\Pi^\epsi_I H^{\epsi}_{\ro{BF}} \Pi^\epsi_I$ is unitarily equivalent to an effective Hamiltonian $H_{\rm eff}$ given by Peierls substitution on some simple reference space~$\mathcal{H}_{\rm ref}$.

Let us  quickly summarize how this is achieved  in the case $B_{0}\equiv0$ in \cite{PST03b,Te03}.  
The smoothness of  $H (k,r)$ 
and the gap condition imply the smoothness of the spectral projection $P_I(k-A(r)) $. 
In particular, $P_I(k-A(r))$ has constant rank $m\in\N$. It is thus natural to choose $\mathcal{H}_{\rm ref}  $ as the $\C^m$-valued functions over the torus $\mathbb{T}^*= \R^2/\Gamma^*$, i.e.\ $\mathcal{H}_{\rm ref} = L^2(\mathbb{T}^*, \C^m)$ .
As in the case of $\Pi^\epsi_I$, the unitary map $U^{\eps}:\Pi^{\eps}_I\mathcal{H}_{\tau}\to\mathcal{H}_{\mathrm{ref}}$ is constructed   perturbatively order by order as the quantization of a semiclassical  symbol $u(k,r) \asymp \sum_{j=0}^\infty \epsi^j u_j(k,r)$. The starting point of the construction is 
a unitary map $u_0(k,r): P_I(k-A(r)) \mathcal{H}_{\rm f} \to \C^m$ that is smooth and  
  right-$\tau$-equivariant,
\[
u_0(k-\gamma^*,r) = u_0(k,r) \tau(\gamma^*)^{-1}\qquad\mbox{for all $k\in \R^2$ and $\gamma^*\in\Gamma^*$}\,.
\]
In geometric terms this means that we seek a $U(m)$-bundle-isomorphism between the Bloch bundle $\bu_{\rm Bl}$ and the trivial bundle over the torus $\mathbb{T}^*$ with fiber $\C^m$.
But such an isomorphism exists if and only if the Bloch bundle is trivial. It was shown in \cite{HS89} for the case $m=1$ and   in \cite{P07} also for $m\geq 1$, that in the   case  $B_0= 0$ time-reversal invariance implies that the Bloch bundle associated to any isolated family of Bloch bands is indeed trivial and hence an appropriate $u_0$ always exists. 

However, when $B_0\not= 0$, then $H_{\rm MB}$ is no longer time-reversal invariant
and the Bloch bundle is in general a non-trivial vector bundle over the torus. Indeed, its non-vanishing Chern numbers are closely related to the quantum Hall effect as was first discovered in the seminal paper by Thouless et al.
\cite{TKNN82}. The non-triviality of magnetic Bloch bundles is  
  the main obstruction for defining Peierls substitution for magnetic Bloch bands in any straightforward way.

Let us start with a rough sketch of our strategy for overcoming this obstruction. 
Our reference space $\mathcal{H}_{\rm ref}=\mathcal{H}_\alpha$ now contains sections of a  non-trivial vector bundle $\bu_\alpha$  over $\mathbb{T}^*$ with typical fiber~$\C^m$ that is isomorphic to the Bloch bundle $\bu_{\rm Bl}$.
According to a result of Panati \cite{P07},  $\bu_\alpha$  is uniquely characterized, up to isomorphisms, by its rank $m\in\N$ and its Chern number $\theta\in\Z$. 
Of course we could just glue together local trivializations of $\bu_{\rm Bl}$  by suitable transition functions in order to construct such a bundle $\bu_\alpha$. However, for the definition of the map $U^\epsi:\Pi^\epsi_I\mathcal{H}_\tau \to \mathcal{H}_\alpha$ and for the construction of an appropriate pseudodifferential calculus 
on $\mathcal{H}_\alpha$, it will be essential to have an explicit characterization of $\bu_\alpha$ with certain additional properties. 
To this end we first  explicitly define a global trivialization of 
  the extended Bloch bundle   given by
\begin{equation}\label{blprime def}
\bu_{\mathrm{Bl}}':=\{(k,\varphi)\in \R^{2} \times \mathcal{H}_{\mathrm{f}}\,|\,  \varphi \in P_I(k)\mathcal{H}_{\mathrm{f}} \}
\end{equation}
over the contractible base space $\R^2$, i.e.\ an orthonormal basis
$(\varphi_1(k), \ldots,\varphi_m(k))$  of $ P(k)\mathcal{H}_{\mathrm{f}}$ depending smoothly on $k\in \R^{2}$. For this we use the parallel transport with respect to 
  the Berry connection $\nabla_k^{\mathrm{B}}=P_I(k)\,\nabla_k \,P_I(k) + P_I^\perp(k)\,\nabla_k \,P_I^\perp (k)$.
 Then $\bu_\alpha :=  (\R^{2}\times \C^m)/_{\displaystyle {\sim}_\alpha  }$
is defined in terms of the ``transition function'' $\alpha:\R^2/\Gamma^*\times \Gamma^*\to \mathcal{L}(\C^m)$ defined by 
$\varphi(k-\gamma^*) =: \alpha(k,\gamma^*)\tau(\gamma^*)\varphi(k)$.
But the functions $\varphi_j(k)$ are not $\tau$-equivariant and their derivatives of order $n$ grow  like $|k|^{n}$. Thus they can not be  used directly to define a symbol of the form $u_0(k,r)_{ij} = |e_i\rangle\langle  \varphi_j(k-A(r))|$. 
However, they do give the starting point for the perturbative construction of a unitary 
$U^\epsi_1: \Pi^\epsi_I\mathcal{H}_\tau \to P_I\mathcal{H}_\tau$ by setting
$u_0(k,r)_{ij} := |\varphi_i(k)\rangle \langle \varphi_j(k-A(r))|$, which is a good $\tau$-equivariant symbol. 
From the frame $(\varphi_1(k), \ldots,\varphi_m(k))$ we also get a bundle isomorphism between $\bu_{\rm Bl}$ and $\bu_\alpha$, i.e.\  a unitary map 
\[
U_\alpha:P_I\mathcal{H}_\tau \to \mathcal{H}_\alpha\,,\quad \varphi(k) \mapsto (U_\alpha\varphi)_j (k) := \langle \varphi_j(k),\varphi(k)\rangle_{\mathcal{H}_{\rm f}} \,,
\]
where $P_I\mathcal{H}_\tau = \{ f\in \mathcal{H}_\tau\,|\, f(k) = P_I(k)f(k)\}$ contains the $L^2$-sections of the Bloch bundle. 
But $U_\alpha$ is not a pseudodifferential operator
and thus it is not clear a priori if 
\[
H_{\rm eff} := U_\alpha \,U^\epsi_1\,  \Pi^\epsi_I \,{\rm Op}^{\tau}\hspace{-1pt}(H)\, \Pi^\epsi_I\, U^{\epsi\,*}_1 \,U_\alpha^*
\]
is a pseudodifferential operator and how its principal symbol looks like. This problem will be solved by introducing a Weyl quantization adapted to the geometry of the Bloch bundle, for which the action of $U_\alpha$ is explicit. 

 After this rough sketch of the general strategy let us 
 start  with the construction of the frame $(\varphi_1(k), \ldots,\varphi_m(k))$. 
 For this we need a lemma on the properties of the Berry connection.

\begin{lem}\label{lemma tau equiv varphi}
On the trivial bundle $ \R^2\times \Hi_{\rm f}$  
 the Berry connection 
 \[
 \nabla^{\mathrm{B}}_k:=P_I(k)\,\nabla_k \,P_I(k) + P_I^\perp(k)\,\nabla_k \,P_I^\perp (k)\,
 \]
  is a metric connection.\\[2mm] For arbitrary $x, y\in \R^{2}$ let $t^{\mathrm{B}} (x,y)$ be the parallel transport with respect
to the Berry connection along the straight line from $y$ to $x$. Then $t^{\mathrm{B}} (x,y)\in \mathcal{L}(\Hi_{\rm f})$ is unitary, satisfies
\begin{equation}\label{berryconP}
t^{\mathrm{B}} (x,y) = P_I(x)t^{\mathrm{B}} (x,y)P_I(y) + P_I^\perp(x)t^{\mathrm{B}} (x,y)P_I^\perp(y)\,,
\end{equation}
 and is $\tau$-equivariant, i.e.\
\begin{equation}\label{tau-equivarianz tB}
t^{\mathrm{B}}(x-\gamma^{*},y-\gamma^{*})=\tau(\gamma^{*})\,t^{\mathrm{B}}(x,y)\,\tau(\gamma^{*})^{-1}.
\end{equation}
\end{lem}
\begin{pro} Let  $\psi,\phi:\R^2\to \Hi_{\rm f}$ be smooth functions, then  a simple computation yields
\[
\nabla  \langle \psi(k),\phi(k)\rangle_{\Hi_{\rm f}} =     \langle\nabla^{\rm B}   \psi(k),\phi(k)\rangle_{\Hi_{\rm f}} +   \langle  \psi(k),\nabla^{\rm B} \phi(k)\rangle_{\Hi_{\rm f}}\,,
\] 
showing that $\nabla^{\rm B}$ is metric.
As a consequence, $t^{\mathrm{B}} (x,y)\in \mathcal{L}(\Hi_{\rm f})$ is unitary. Let $x(s) := y + s(x-y)$, $s\in[0,1]$, be the straight line from $y$ to $x$.
Then $t^{\rm B}(x(s),y)=:t^B(s) $ is the unique solution of 
\begin{equation}\label{diff1}
\frac{\D}{\D s} t^B(s) = [ (x-y)\cdot\nabla  P_I(x(s)), P_I(x(s)) ]\,t^B(s) \quad\mbox{with}\quad t^B(0)={\bf 1}_{\mathcal{H}_{\rm f}}\,.
\end{equation}
From this and $\nabla P_I =P_I (\nabla P_I)P_I^\perp+P_I^\perp(\nabla P_I)P_I$ one easily computes that
\[
\frac{\D}{\D s} \left(t^{B }(s)^* P_I(x(s)) t^B(s)\right) =0 \,,
\]
which implies $t^{B }(s)^* P_I(x(s)) t^B(s)=P_I(y)$ for all $s\in[0,1]$ and thus \eqref{berryconP}.
Now
$t^B(x(s)-\gamma^*,y-\gamma^*):=\tilde  t^B (s)$ is the unique solution of 
\begin{equation}\label{diff2}
\frac{\D}{\D s} \tilde t^B(s) = [(x-y)\cdot\nabla P_I(x(s)-\gamma^*), P_I(x(s)-\gamma^*)]\,\tilde t^B(s) \quad\mbox{with}\quad\tilde t^B(0)={\bf 1}_{\mathcal{H}_{\rm f}}\,.
\end{equation}
Thus the $\tau$-equivariance of $t^{\mathrm{B}} (x,y)$
 follows from comparing \eqref{diff1} and \eqref{diff2} and using the $\tau$-equivariance of the projection $P_I(k)$.
\end{pro}

\begin{prop} \label{varphi} Let $\{E_{n}(k)\}_{n\in I}$  be an isolated family of Bloch bands with $|I|=m$.
There are functions $\varphi_j \in C^{\infty}(\R^{2},\mathcal{H}_{\mathrm{f}})$, $j=1,\ldots,m$,  such that $(\varphi_1(k), \ldots,\varphi_m(k))$ is an orthonormal basis of $ P_I(k)\mathcal{H}_{\mathrm{f}}$ for all $k\in \R^{2}$ and having the following property:\\ There is a  function $\alpha:\R/\Z  \to \mathcal{L}(\C^m)$ taking values in the unitary matrices, such that 
\[
\varphi(k-\gamma^{*})= \alpha\left( \kappa_2\right)^{n_1} \tau(\gamma^{*})\varphi(k) 
\]
for all
$
 \gamma^{*}=:n_1\gamma^*_1+n_2 \gamma^*_2  \in \Gamma^{*}$, $  k  \in\R^2
$
and $\kappa_2:= \frac{\langle k,\gamma_2\rangle}{2\pi}$.
If the rank $m$ of the Bloch bundle is one, then $\varphi=\varphi_1$ can be chosen such that  \begin{equation}\label{alphacanon}
\alpha(\kappa_2)=\E^{- \I 2\pi \theta \kappa_2} =\E^{- \I   \theta \langle k,\gamma_2\rangle }\,,\end{equation}
where $\theta\in\Z$ is the Chern number of the Bloch bundle.
\end{prop}

\begin{pro}  
Note that if the Bloch bundle  is trivial, then  any trivializing frame  $(\varphi_j(k))_{j=1,\ldots,m}$ would do the job and $\alpha\equiv {\bf 1}_{m\times m}$. In general, we    construct a trivializing frame of the extended Bloch bundle $\bu_{\mathrm{Bl}}'$ (see  \eqref{blprime def})    by using the parallel 
transport with respect to the Berry connection.

 Throughout this proof we use instead of cartesian coordinates the coordinates $\kappa_j:= \frac{\langle k,\gamma_j\rangle}{2\pi}$, i.e.\ $k  = \kappa_1\gamma_1^* + \kappa_2\gamma^*_2$. In particular, we identify also $\gamma^* = (n_1,n_2) \in\Gamma^*$ with $(n_1,n_2)\in\Z^2$.

Let $\kappa_2\mapsto (h_1(\kappa_2),\ldots,h_m(\kappa_2))$ be a smooth  $\tau_2$-equivariant   orthonormal frame of  
$\bu_{\rm Bl}'|_{\kappa_1=0}$, i.e.\ $h_j(\kappa_2 - n_2) = \tau((0,n_2))h_j(\kappa_2)$ and $(h_1(\kappa_2),\ldots,h_m(\kappa_2))$ is an orthonormal basis of \linebreak
$P_I((0,\kappa_2))\mathcal{H}_{\mathrm{f}}$.  Since every complex vector bundle over the circle is trivial, such a frame always exists. 
Now we define a global frame of $E'_{\rm Bl}$ by parallel transport of $h$ along the $\gamma_1^*$-direction, i.e.\
\[
\tilde{\varphi}_j(\kappa_{1},\kappa_{2}):= t^{\mathrm{B}}((\kappa_{1},\kappa_{2}),(0,\kappa_{2}))\,h_j (\kappa_{2})\,.
\]
By Lemma~\ref{lemma tau equiv varphi}, the functions $\tilde{\varphi}_j:\R^2\to\mathcal{H}_{\rm f}$ are   smooth and $(\tilde{\varphi}_1(k),\ldots,\tilde{\varphi}_m(k))$ is an orthonormal basis of $P_I(k)\mathcal{H}_{\mathrm{f}}$   for all
$k\in \mathbb{R}^{2}$.
Since    $\tau(\gamma^*): {\rm Ran} P_I(k) \to {\rm Ran} P_I(k+\gamma^*)$ is unitary  for all $k\in\R^2$, we have that
 \begin{equation}\label{def alpha}
 \tilde{\varphi}_j(k-\gamma^{*})=:\sum_{i=1}^m  {\tilde{\alpha}_{ji}(k,\gamma^{*})}\tau(\gamma^{*})\tilde{\varphi}_i(k)
\end{equation} 
with a unitary $m\times m$-matrix $\tilde \alpha(k,  \gamma^* ) = (\tilde{\alpha}_{ji}(k,\gamma^{*}))_{j,i=1,\ldots ,m}$.  
The $\tau$-equivariance of $h$ implies 
\[
\tilde\alpha((0,\kappa_2),(0,n_2))={\bf 1}_{m\times m} \quad \mbox{ for all $\kappa_2\in\R$ and $n_2\in\Z$  } \,.
\]
From the $\tau$-equivariance \eqref{tau-equivarianz tB} of the parallel transport  this implies also 
\begin{equation}\label{alphaper}
  \tilde \alpha(k, (0,n_2))={\bf 1}_{m\times m}\quad \mbox{ for all $k\in\R^2$ and $n_2\in\Z$ }\,,
\end{equation}
since  
\begin{eqnarray*}\lefteqn{
t^{\mathrm{B}}((\kappa_1,\kappa_2-n_2),(0,\kappa_{2}-n_2))\; \tau((0,n_2))\;t^{\mathrm{B}}((0,\kappa_2),(\kappa_1,\kappa_{2})) =}\\
&=& \tau((0,n_2))\;
t^{\mathrm{B}}((\kappa_1,\kappa_2 ),(0,\kappa_{2} ))\; \tau((0,n_2))^{-1}\; \tau((0,n_2))\;t^{\mathrm{B}}((0,\kappa_2),(\kappa_1,\kappa_{2})) \\
&=& \tau((0,n_2))\,.
\end{eqnarray*}
From the definition \eqref{def alpha}  it follows that $\tilde \alpha$ satisfies the cocycle condition 
\begin{equation}\label{cocycle}
\tilde \alpha(k-\tilde \gamma^*,\gamma^*)\,\tilde\alpha(k,\tilde\gamma^*) = \tilde\alpha(k,\gamma^*+\tilde\gamma^*)\quad\mbox{ for all $k\in\R^2$ and $\gamma^*,\tilde\gamma^*\in\Gamma^*$,}
\end{equation}
which for $\gamma^* = (0,n_2)$ and $\tilde\gamma^*=(n_1,0)$ together with \eqref{alphaper} implies 
\[
 \tilde \alpha(k,(n_1,0))=\tilde\alpha(k, (n_1,n_2))  \quad \mbox{ for all $k\in\R^2$ and $n_1,n_2\in\Z$ }\,.
\]
Hence $\tilde\alpha$ does not depend on $n_2$ and we write $\tilde\alpha(k,n_1)$ in the following. 
But then the cocycle condition \eqref{cocycle} with $\gamma^* = (n_1,0)$ and $\tilde\gamma^*=(0,n_2)$  implies 
\[
\tilde\alpha((\kappa_1,\kappa_2-  n_2),n_1)\,\tilde\alpha((\kappa_1,\kappa_2), 0) = \tilde\alpha((\kappa_1,\kappa_2),n_1 )\,,
\]
and thus periodicity   of  $\tilde \alpha$ as a function of $\kappa_2$.

Next we introduce the $m\times m$-matrix-valued connection coefficients of the Berry connection as
\[
 \begin{pmatrix} \tilde{\mathcal{A}}^1_{ji }(k)\\ \tilde{\mathcal{A}}^2_{ji}(k)  \end{pmatrix}
 := - \frac{\I}{2\pi}
  \begin{pmatrix} \left< \tilde{\varphi}_i(k),\partial  _{\kappa_{1}} \tilde{\varphi}_j(k)\right>_{\mathcal{H}_{\mathrm{f}}}\\\left< \tilde{\varphi}_i(k),\partial _{\kappa_{2}} \tilde{\varphi}_j(k)\right>_{\mathcal{H}_{\mathrm{f}}} \end{pmatrix} =
   \begin{pmatrix}0\\ \tilde{\mathcal{A}}^2_{ji}(k) \end{pmatrix}\,,
\]
where $ \tilde{\mathcal{A}}^1_{ji}(k)=0$ because the $\tilde \varphi_i$ are parallel along the $\gamma^*_1$-direction.    From \eqref{alphaper} we infer  that $\tilde{\mathcal{A}}^2$ is periodic in the $\gamma^*_2$-direction, i.e.\ that $\tilde{\mathcal{A}}^2( \kappa_1,\kappa_2+n_2)=\tilde{\mathcal{A}}^2(\kappa_1, \kappa_2  )$ for all $k\in \R^2$ and $n_2\in\Z$.

If we differentiate both sides of \eqref{def alpha} with respect to $\kappa_\ell$ and then project on $\tilde\varphi_s(k-\gamma^*)$, we obtain  
\begin{eqnarray*}
2\pi \I \,\tilde{\mathcal{A}}^\ell_{js}(k-\gamma^*)
&=&   \sum_{i=1}^m\big( \langle \tilde\varphi_s(k-\gamma^*),\partial_{\kappa_\ell} \tilde \alpha_{ji}(k,n_1) \tau(\gamma^*)\tilde\varphi_i(k)
  + \tilde\alpha_{ji}(k,n_1)   \tau(\gamma^*)
\partial_{\kappa_\ell} \tilde\varphi_i(k)
\rangle \big)
\\& = & \sum_{i=1}^m \partial_{\kappa_\ell} \tilde \alpha_{ji}(k,n_1) \, \overline{\tilde\alpha_{si}(k,n_1)} + 2\pi \I \sum_{i,n=1}^m \tilde \alpha_{ji}(k,n_1) \,  \tilde{\mathcal{A}}^\ell_{in}(k)\, \overline{\tilde\alpha_{sn}(k,n_1)}\,.
\end{eqnarray*}
Since $ \tilde{\mathcal{A}}^1_{ji}(k)=0$, the matrix $\tilde\alpha(k,n_1)$ is independent of $\kappa_1$ and satisfies the  linear first order ODE
\begin{equation}\label{alphaODE}
\partial_{\kappa_2} \tilde \alpha (\kappa_2,n_1) = 2\pi\I\, \left(\tilde{\mathcal{A}}^2 (0,\kappa_2)
\,\tilde \alpha (\kappa_2,n_1) \,
-\tilde \alpha (\kappa_2,n_1) \,\tilde{\mathcal{A}}^2 (n_1,\kappa_2)
\right) \,.
\end{equation}
Since
 $\tilde\alpha(\kappa_2,\cdot):\Z\to \mathcal{L}(\C^m)$ is a group homomorphism for every $\kappa_2\in\R/\Z$, we can put  $\tilde\alpha(\kappa_2,n_1) = \alpha(\kappa_2)^{n_1}$ with $\alpha(\kappa_2):=\tilde \alpha(\kappa_2,1)$. This proves the statement of the lemma for the case $m>1$ by setting $\varphi:=\tilde\varphi$.

For $m=1$ we evaluate the solution of \eqref{alphaODE} in order to obtain an explicit expression for $\alpha$,
\[
\tilde\alpha ( \kappa_2,1) = \exp\left( 2\pi\I \int_0^{\kappa_2}\D s \left( \tilde{\mathcal{A}}^2( 0,s ) -  \tilde{\mathcal{A}}^2( 1,s )
\right)
\right)\,.
\]
Introducing  the curvature of the Berry connection
\[
\Omega(k)  = \frac{|M^*|}{2\pi } \,\partial_{\kappa_1} \,\tilde{\mathcal{A}}^{2}(k) \,,
\]
   by Stokes' theorem we have  
\[
 2\pi\int_{0}^{\kappa_{2}}(\tilde{\mathcal{A}}^{2}(1,s)-\tilde{\mathcal{A}}^{2}(0,s))\ro{d}s
=  \frac{4\pi^2}{|M^*|} \int_{0}^{\kappa_{2}} \int_{0}^{1} \Omega(p,s)\,\mathrm{d}p \,\ro{d}s
 =:     \overline{\Omega}(\kappa_{2})
\]
and thus
\[
\tilde\alpha(\kappa_2,1) = \E^{-\I      \overline{\Omega}(\kappa_{2})  }  \,.
\]
To obtain the simpler form claimed in the Lemma, we  put
$$\varphi(k):= \E^{\I   \kappa_{1}\left( 2\pi  \kappa_{2} \theta-   \overline{\Omega}(\kappa_{2}) \right)}\tilde{\varphi}(k),$$
where $\theta := \frac{\overline{\Omega}(1)}{2\pi}$ is the Chern number of the Bloch bundle. Hence
$$
\varphi(k-\gamma^*) = \E^{-\I2\pi \theta  \kappa_{2}n_1} \tau(\gamma^*)\varphi(k) \,. $$ 
\end{pro}

\begin{prop}\label{u1epsilon}
Let Assumptions~\ref{A_1: A, Phi, V_Gamma} and \ref{A_2: B0rational} hold with $A^{(1)}=0$ and 
let $\{E_{n}(k)\}_{n\in I}$  be an iso\-lated family of Bloch bands.
 Then there exists a unitary ope\-rator $U^{\eps}_{1}\in\mathcal{L}( \mathcal{H}_{\tau} )$ such that
\begin{equation*}\label{UpiU=PI0}
U^{\eps}_{1} \Pi^{\varepsilon}_I U^{\varepsilon*}_{1}=P_I
\end{equation*}
and $U^{\eps}_{1}={\rm Op}^{\tau}\hspace{-1pt}(u)+\mathcal{O}_{0}(\varepsilon^{\infty})$, where $u \asymp \sum _{j\geq 0} \varepsilon^{j}u_{j}$ belongs to $S^{1}_{\tau}(\varepsilon, \mathcal{L}(\mathcal{H}_{\mathrm{f}}))$ and
has the $\tau$-equivariant principal symbol $u_{0}(k,r)= \sum_{i=1}^m \left|\varphi_i(k)\right>\left<\varphi_i(k-A(r))\right| +u_{0}^{\perp}(k,r)$.
\end{prop}

\begin{pro}  We only need to show that a $\tau$-equivariant principal symbol $u_{0}(k,r)$ of the form claimed above exists. Then 
the proof works line by line as the proof of Proposition~5.18 in \cite{Te03}, see also \cite{PST03b}. 
However, according to Lemma~\ref{lemma tau equiv varphi},
\[
u_0(k,r):= t^{\rm B}(k,k-A(r)) = t^{\rm B}((\kappa_1,\kappa_2),(\kappa_1-A_1(r),\kappa_2))
\]
 is $\tau$-equivariant and has the desired form. Here we use the choice of gauge $\gamma_2\cdot A(r)=0$ and write as before $A(r)= A_1(r)\gamma_1^*$.
Note that at this point we have to assume $A^{(1)}\equiv 0$, because otherwise the $\kappa_2$-derivatives of $u_0$ would become unbounded functions of $r$ and $u_0\notin S^w_\tau$ for all order functions $w$.

 \end{pro}

\begin{defi}\label{uthetadef}
 Using the matrix-valued function $\alpha$ constructed in Proposition~\ref{varphi}, we define
 \[
 \mathcal{H}_\alpha :=  \left\{ f \in L^{2}_{\mathrm{loc}}(\R ^{2},\C^m)\,|\,f(k-\gamma^{*})=\alpha(\kappa_2)^{-n_1} f(k)\, \mbox{ for all } \,k \in \R^{2} \,,\; \gamma^{*} 
\in \Gamma^{*}\right\}
\]
 with inner product $\langle f,g \rangle_{\mathcal{H}_{\alpha}}= \int _{M^{*}}\mathrm{d}k\,\langle{f(k)},g(k)\rangle_{\C^m}$.\\[2mm]
 Using the orthonormal frame $(\varphi_1(k), \ldots,\varphi_m(k))$ constructed in Proposition~\ref{varphi}, we define the unitary maps
 \[
 U_\alpha(k) : P_I(k) \Hi_{\rm f} \to \C^m\,,\quad f\mapsto  (U_\alpha (k)  f)_i := \left<\varphi_i(k),f\right>_{\mathcal{H}_{\mathrm{f}}}
 \]
 and
  \[
 U_{\alpha} : P_I \mathcal{H}_{\tau} \rightarrow \mathcal{H}_{\alpha}\,,\quad f \mapsto (U_\alpha f) (k)_i := \left<\varphi_i(k),f(k)\right>_{\mathcal{H}_{\mathrm{f}}}\,.
 \]
\end{defi}

In the same way that $P_I\mathcal{H}_\tau$ is the space of $L^2$-sections of the Bloch bundle $\bu_{\rm Bl}$, the space $\mathcal{H}_\alpha$ is the space of $L^2$-sections of a bundle $\bu_\alpha$.  
\begin{defi}
Let
\begin{equation}\label{E_theta}
\bu_\alpha :=  (\R^{2}\times \C^m)/_{\displaystyle {\sim}_\alpha  },
\end{equation}
where
$$  (k,\lambda) \sim_\alpha (k',\lambda')\quad :\Leftrightarrow \quad k'=k-\gamma^{*}  \mbox{ and }  \lambda '=\alpha(\kappa_2)^{-n_1}\lambda \quad\mbox{for some $\gamma^*=(n_1,n_2)\in\Gamma^*$}\,.$$
On sections of $\bu_\alpha$ we define  the connection $\nabla^\alpha := U_\alpha \nabla^{\rm B} U_{\alpha}^*$.
\end{defi}
It was shown by  Panati \cite{P07} that even for $m>1$ the bundle $\bu_\alpha$ is, up to isomorphisms, uniquely determined by its Chern number
\[
\theta := \frac{1}{2\pi} \int_{M^*}{\rm tr}(\Omega(k))\,\D k \,.
\]
However, we use a canonical form for $\alpha$ only in the case $m=1$, where a canonical choice is~\eqref{alphacanon}.

\section{The effective Hamiltonian  as a  pseudodifferential operator}

Combining the unitary maps $U^\epsi_1: \Pi^\epsi_I \mathcal{H}_\tau\to P_I \mathcal{H}_\tau$
and $U_\alpha:  P_I \mathcal{H}_\tau \to \mathcal{H}_\alpha$ into 
\[
U^\epsi:  \Pi^\epsi_I \mathcal{H}_\tau \to \mathcal{H}_\alpha\,,\quad U^\epsi : = U_\alpha U^\epsi_1\,,
\]
we find that the block $ \Pi_I^{\eps}  H^{\eps}_{\ro{BF}}\Pi_I^{\eps}$ of $H^{\eps}_{\ro{BF}}$ is unitarily equivalent to the effective Hamiltonian 
$$H^{\mathrm{eff}}_I: = U^{\eps} \Pi_I^{\eps}H^{\eps}_{\ro{BF}} \Pi_I^{\eps} U^{\eps *}$$
acting on the space $\mathcal{H}_\alpha$ of $L^2$-sections of $\bu_\alpha$. 
The remaining problem is to compute    explicitly  an asymptotic expansion of $H^{\mathrm{eff}}_I$ in powers of $\epsi$, where the leading order term should be given by Peierls substitution,
\[
H^{\mathrm{eff}}_I =  E_I(k-A(\I\epsi\nabla_k^\alpha))  +\Phi(\I\epsi\nabla_k^\alpha ) + \Or(\epsi)
\]
with
\[
E_I(k)_{ij} = \langle \varphi_i(k) , H_{\rm per}(k) \varphi_j(k)\rangle\,.
\]
Note that $\nabla^\alpha$ is the only natural connection on sections of $\bu_\alpha$, as the flat connection, used implicitly for Peierls substitution in the non-magnetic case, is not at our disposal. 
It will be a considerable effort in itself to properly define the pseudo\-differential operator $E_I(k-A(\I\epsi\nabla_k^\alpha))  +\Phi(\I\epsi\nabla_k^\alpha )$ as an operator on $\mathcal{H}_\alpha$.

In the non-magnetic case the problem of expanding $H_{\rm eff}$ is much simpler.  Then not only the Hamiltonian $H^{\eps}_{\ro{BF}}={\rm Op}^{\tau}\hspace{-1pt}(H)$ and the projection $\Pi_I^{\eps}=  {\rm Op}^{\tau}\hspace{-1pt}(\pi)+\mathcal{O}(\eps^{\infty})$ 
are $\mathcal{O}(\eps^{\infty})$-close
to pseudodifferential operators, but also 
 the intertwining unitary $U^{\eps}={\rm Op}^{\tau}\hspace{-1pt}(u) +\mathcal{O}(\eps^{\infty})$. Moreover,  $\mathcal{H}_\alpha$ contains periodic functions and   $H^{\mathrm{eff}}_I$ is close to a semiclassical pseudodifferential operator $ {h}^{\rm eff}_I(k,\I\epsi\nabla_k)$ with an asymptotic expansion of its symbol computable using the Moyal product,
\begin{eqnarray*}
H^{\mathrm{eff}}_I&=&U^{\eps}\Pi^{\eps}_IH^{\eps}_{\mathrm{BF}}\Pi^{\eps}_IU^{\eps *}= {\rm Op}^{\tau}\hspace{-1pt}(u) \,{\rm Op}^{\tau}\hspace{-1pt}(\pi)\,{\rm Op}^{\tau}\hspace{-1pt}(H)\,{\rm Op}^{\tau}\hspace{-1pt}(\pi)\,{\rm Op}^{\tau}\hspace{-1pt}(u^*) +\mathcal{O}(\eps^{\infty})\\&=&\mathrm{Op}^\tau(\underbrace{u\sharp \pi\sharp H\sharp \pi \sharp u^{*}}_{=:h_I^{\rm eff}})+\mathcal{O}(\eps^{\infty}) \,.
\end{eqnarray*}
In our magnetic case, however, we cannot proceed like this. Although the operators $\Pi_I^{\eps}$ and $U_{1}^{\eps}$ are again nearly pseudodifferential operators,
this is no longer true for~$U_{\alpha}$. The symbol for this operator would have to be $u_{\alpha}(k,r)=\sum _{i=1}^{m}\la\varphi_i(k)|$, which is in no suitable symbol class because its derivatives of order $n$ grow like $|k|^{n}$.
  So we have to deal with the fact that our effective Hamiltonian
is of the form
$$H^{\mathrm{eff}}_I= U^{\eps} \Pi^{\eps}_I \,{\rm Op}^{\tau}\hspace{-1pt}(H) \,\Pi^{\eps}_I U^{\eps *}=U_{\alpha} P_I\,{\rm Op}^{\tau}\hspace{-1pt}(\mathfrak{h})P_I U_\alpha^{ *}+\mathcal{O}(\eps^{\infty})\,.
$$
Our solution is to replace the $\tau$-quantized operator  ${\rm Op}^{\tau}\hspace{-1pt}(\mathfrak{h})=\mathfrak{h}(k,\I\epsi\nabla_k^\tau)$
by a ``Berry quantized'' operator ${\rm Op}^{\rm B}\hspace{-1pt} (h ) = h (k, \I\epsi\nabla_k^{\rm B})$ (c.f.\ \eqref{berryQdef}) with a modified  symbol $h $. Because of the unitary equivalence $\nabla^\alpha  = U_\alpha \nabla^{\rm B} U_{\alpha}^*$, one expects and we will show that $U_\alpha h (k, \I\epsi\nabla_k^{\rm B}) U_{\alpha }^* =  h^{\rm eff}_I(k, \I\epsi\nabla_k^\alpha) $
with $h^{\rm eff}_I(k,r)_{ij} := \langle \varphi_i(k), h (k,r) \varphi_j(k)\rangle$.
We postpone the detailed definitions of the new quantizations and the proofs of their relevant properties to Section~\ref{quantizations}. In a nutshell the quantization maps are defined as follows.
\begin{itemize}
 \item For $h\in S_\tau^1(\epsi, \mathcal{L}( \mathcal{H}_{\rm f}))$ we put ${\rm Op}^{\rm B}\hspace{-1pt} (h )  = h(k,\I\epsi\nabla^{\rm B}_k) \text{ acting on }  \mathcal{H}_{\tau}$
 \item For $h\in S_{\alpha}(\epsi, \mathcal{L}( \C^m))$ (c.f.\ Definition~\ref{Salpha}) we put ${\rm Op}^{\alpha}\hspace{-1pt} (h )   = h(k,\I\epsi\nabla^{\alpha}_k) \text{ acting on }  \mathcal{H}_{\alpha}$
 \item For $m=1$ and $\Gamma^*$-periodic $h\in S^1 (\epsi, \mathcal{L}( \C ))$ we put ${\rm Op}^{\theta}\hspace{-1pt} (h )  = h(k,\I\epsi\nabla^{\theta}_k)$\\   acting on $\mathcal{H}_{\alpha}$, where $\nabla_k^{\theta} := \nabla_{k}+   \tfrac{\I \theta}{2\pi}  \langle k,\gamma_1 \rangle \,\gamma_2$.
\end{itemize}
The last quantization will only be used for the case $m=1$ in order to obtain an explicit expression for $H^{\mathrm{eff}}_I$.
Note that changing  the connection from $\nabla^\alpha$ to $\nabla^\theta$ makes the quantization rule   independent of $\varphi_1$. Moreover, $\nabla^\theta$ is canonical in the sense that its curvature tensor
$R^\theta (X,Y) = \tfrac{ \I \theta |M|}{2\pi} (X_1Y_2-X_2Y_1)$ is constant. 

All in all, the steps leading to a representation of the effective Hamiltonian $H^{\rm eff}_I$ as a pseudodifferential operator are 
\begin{eqnarray*}
H^{\mathrm{eff}}_I&:=& U^{\eps}\Pi_I^{\eps}H^{\eps}_{\mathrm{BF}}\Pi_I^{\eps}U^{\eps *}=U^{\eps} \Pi^{\eps}_I \,{\rm Op}^{\tau}\hspace{-1pt}(H) \,\Pi^{\eps}_I U^{\eps *}=U_{\alpha} P_I\,{\rm Op}^{\tau}\hspace{-1pt}(\mathfrak{h})P_I U_\alpha^{ *}+\mathcal{O}(\eps^{\infty})\\&=&U_{\alpha}P_I \,{\rm Op}^{\rm B}\hspace{-1pt} (h ) \,P_I U_{\alpha }^*+\mathcal{O}(\eps^{\infty})={\rm Op}^{\alpha}\hspace{-1pt} (h^{\rm eff}_I)  +\mathcal{O}(\eps^{\infty})\,.
\end{eqnarray*}
In the following theorem we collect our main results.

\begin{thm} \label{5.11}
Let Assumptions~\ref{A_1: A, Phi, V_Gamma} and \ref{A_2: B0rational} hold with $A^{(1)}=0$ and 
let $\{E_{n}(k)\}_{n\in I}$  be an isolated family of Bloch bands.
Then there exist
an orthogonal projection $\Pi^{\eps}_I \in \mathcal{L}(\mathcal{H}_{\tau})$
and a unitary map $U^{\eps} \in \mathcal{L}(\Pi^{\varepsilon}_I\mathcal{H}_{\tau},\mathcal{H}_{\alpha})$ 
such that 
\begin{equation}\label{state1}
\left\| [H_{\mathrm{BF}}^{\eps},\Pi^{\eps}_I] \right\| _{\mathcal{L}(\mathcal{H}_{\tau})}= \mathcal{O}(\eps^{\infty})\,.
\end{equation}
and, with $H^{\mathrm{eff}}_I := U^{\eps}\Pi_I^{\eps}H^{\eps}_{\mathrm{BF}}\Pi_I^{\eps}U^{\eps *}$,
\begin{equation}\label{state2}
\left\| (\E^{-\I H^{\varepsilon}_{\mathrm{BF}}t}-U^{\eps *}\E^{-\I H^{\mathrm{eff}}_I t}U^{\eps})\Pi_I^{\varepsilon}\right\|_{\mathcal{L}(\mathcal{H}_{\tau})} = \mathcal{O}(\varepsilon^{\infty} |t| )\,.
\end{equation}
If $\{E_{n}(k)\}_{n\in I}$ is strictly isolated with gap $d_{\rm g}$ and $\left\| \Phi \right\|_{\infty} < \tfrac{1}{2}d_{\ro{g}}$, then the expressions in \eqref{state1} and \eqref{state2} vanish exactly.\\[2mm]
There is an $\alpha$-equivariant symbol
$h_I^{\rm eff}\in S_{\alpha}(\epsi, \mathcal{L}( \C^m))$ such that 
\begin{equation}\label{claimt1}
\big\| H^{\mathrm{eff}}_I-{\rm Op}^{\alpha}\hspace{-1pt} (h_I^{\rm eff})   \big\| _{\mathcal{L}(\mathcal{H}_{\alpha})}= \mathcal{O}(\eps^{\infty})\,.
\end{equation}
The asymptotic expansion of the symbol $h_I^{\mathrm{eff}}$ can be computed, in principle, to any order in $\epsi$. Its principal symbol is given by 
\[
h_0(k,r)  =  E_I(k-A(r))  +\Phi(r ) {\bf 1}_{m\times m}\,,
\]
where 
\[
E_I(k)_{ij} := \langle \varphi_i(k) , H_{\rm per}(k) \varphi_j(k)\rangle_{\mathcal{H}_{\rm f}}
\]
and 
$(\varphi_1(k), \ldots,\varphi_m(k))$ is the orthonormal frame of the extended Bloch bundle constructed in Proposition~\ref{varphi}. Thus, Peierls substitution is the leading order approximation to the restriction of the Hamiltonian to an isolated family of bands, 
\[
\big\| H^{\mathrm{eff}}_I - {\rm Op}^{\alpha}\hspace{-1pt} (h_0) \big\| _{\mathcal{L}(\mathcal{H}_{\alpha})}= \mathcal{O}(\eps)\,.
\]
\end{thm}

\begin{pro}  The projection $\Pi^\epsi_I$ was constructed in Theorem~\ref{space-adiabatic threorem}.
The unitary $U^\epsi:= U_\alpha U^\epsi_1$ is obtained from $U^\epsi_1$ constructed in 
Corollary~\ref{u1epsilon} and
$U_\alpha$ given in Definition~\ref{uthetadef}. Statement \eqref{state2} follows from \eqref{state1} by standard time-dependent perturbation theory.

Now the operator  $H_0 :=  U_1^{\eps}\Pi_I^{\eps}H^{\eps}_{\mathrm{BF}}\Pi_I^{\eps}U_1^{\eps *}$ is, by construction, asymptotic to the $\tau$-quantiza\-tion of the semiclassical symbol $\mathfrak{h}:=u\sharp \pi \sharp H \sharp \pi \sharp u^{*}\in S^1_\tau(\epsi)$
with principal symbol
\[
\mathfrak{h}_0(k,r)  =  \langle \varphi_i(k-A(r)) ,( H_{\rm per}(k-A(r)) +\Phi(r) )\varphi_j(k-A(r))\rangle_{\mathcal{H}_{\rm f}}\; |\varphi_i(k)\rangle\langle \varphi_j(k)|\,.
\]
As sketched before and as to be shown in Corollary~\ref{tauberry}, one can approximate ${\rm Op}^{\tau}\hspace{-1pt} (\mathfrak{h})$ by the Berry-quantization  ${\rm Op}^{\rm B}\hspace{-1pt} (h) $ of a modified symbol $h$ up to an error of order $\epsi^{\infty}$. More precisely, in Corollary~\ref{tauberry} we show that  there is a sequence of symbols $h_n\in S^1_\tau$ with $h_0 = \mathfrak{h}_0$ such that for any $N\in\N$
\[
\Big\|\sum_{n=0}^N\epsi^n {\rm Op}^{\rm B}\hspace{-1pt} (h_n) - {\rm Op}^{\tau}\hspace{-1pt} (\mathfrak{h})  \Big\| = \Or(\epsi^{N+1})\,.
\]
As we will show in Proposition~\ref{theta berry}, 
the Berry-quantization transforms in an explicit way under the unitary mapping $U_\alpha$ to the reference space $\Hi_\alpha$.
Namely, it holds that $U_\alpha  {\rm Op}^{\rm B}\hspace{-1pt} (h_n)  U_\alpha^*=   {\rm Op}^{\alpha}\hspace{-1pt}  (h_n^{{\rm eff}})$ with
\[
(h_n^{{\rm eff}})_{ij}(k,r) = \langle \varphi_i(k), h_n(k,r) \varphi_j(k)\rangle\,.
\]
Then  \eqref{claimt1} holds for  any resummation $h_I^{\rm eff}$ of the asymptotic series $\sum\epsi^n h_n^{{\rm eff}}$.
\end{pro}

\medskip 
As stated in the theorem, one can compute order by order the asymptotic expansion of 
$h^{\rm eff}_I$ using the explicit expansions of the symbols $\pi$ and $u$ and expanding Moyal products. We now show how to compute the subprincipal symbol $h_1$ in a special case, and for this  we adopt the notation introduced in the proof of Theorem~\ref{5.11}.
According to Corollary~\ref{tauberry} there are two contributions to $h_1$, 
namely
\[
  h_1(k,r) =h_{1,\rm c} + \mathfrak{h}_1 := -\tfrac{\I}{2} 
\left( \nabla_r \mathfrak{h}_0(k,r) \cdot M(k) + M(k) \cdot  \nabla_r \mathfrak{h}_0(k,r)
\right) \;+\; \mathfrak{h}_1\,,
\]
where
\[
M(k) :=  [\nabla P_I(k),P_I(k)]\,.
\]
While one could compute $h_1$ also for general isolated families of bands, this is more cumbersome and the result is rather complicated. We therefore  specialize to the case $m=1$, i.e.\ to a single nondegenerate isolated band $E_n$. Then
\[
\mathfrak{h}_0 (k,r) =\left(  E_n(k - A(r)) + \Phi(r) \right) P_I(k)
\]
 and using the $\varphi$ corresponding to \eqref{alphacanon} we obtain that the Berry connection coefficient $\mathcal{A}_1(k) = -\tfrac{ \I}{2\pi} \langle \varphi_n(k) ,\partial_{\kappa_1} \varphi_n(k)\rangle $
is a periodic function of  $k_2$ and independent of $k_1$. Hence introducing the kinetic momentum $\tilde{k}:=k-A(r)$ and recalling that $A(r)= A_1(r) \gamma_1^*$, we have $\mathcal{A}_1(\tilde k)= \mathcal{A}_1(k)$.
Using this and specializing to the case $\Gamma=\Z^2$ for the moment, 
one finds for the subprincipal symbol of 
\[
\mathfrak{h}=u\sharp \pi \sharp H \sharp \pi \sharp u^{*} = P_I\sharp u \sharp H \sharp u^* \sharp P_I
\]
by the same reasoning as in the proof of Corollary~5.12 in  \cite{Te03} the expression
\begin{eqnarray*}
\mathfrak{h}_{1}(k,r)&=& \bigg( - \mathcal{A}_{1}(\tilde{k})\left(  \partial_{2}E_n(\tilde{k})B(r)-\partial_{r_{1}} E_n(\tilde{k})\right) \\
&&\;+ \left(\mathcal{A}_{2}(k)-\mathcal{A}_{2}(\tilde{k})\right)\left(\partial_{2}\Phi(r)-\partial_{1}E_n(\tilde{k})B(r)\right)\\
&&\;+B(r) \;\ro{Re} \left(\I \langle\partial_{1}\varphi_n(\tilde{k}),(H_{\mathrm{per}}-E_n)(\tilde{k})\partial_{2}\varphi_n(\tilde{k})\rangle_{\mathcal{H}_{\mathrm{f}}}\right) \bigg)P_I(k)\\
&&\;- \tfrac{\I}{2} \nabla_{r}\left(E_n(\tilde{k})+\Phi(r)\right) M(k),
\end{eqnarray*}
where $\tilde{k}:=k-A(r)$ and $B= {\rm curl}A = \partial_2 A_1$.
Using  $P_I(k) \nabla P_I(k) P_I(k) =0$ the last term in $\mathfrak{h}_1$ cancels exactly $h_{1,\rm c}$ in $h_1$ and we find
\begin{eqnarray*}
h^{\rm eff}_1 (k,r) &=& \langle \varphi_n(k), h_1(k,r) \varphi_n(k)\rangle\\ &=& - \mathcal{A}_{1}(\tilde{k})\left(  \partial_{2}E_n(\tilde{k})B(r)-\partial_{r_{1}} E_n(\tilde{k})\right) \\
&&\;+ \left(\mathcal{A}_{2}(k)-\mathcal{A}_{2}(\tilde{k})\right)\left(\partial_{2}\Phi(r)-\partial_{1}E_n(\tilde{k})B(r)\right)\\
&&\;+B(r) \;\mathcal{M}(\tilde k) \,,
\end{eqnarray*}
with 
\[
\mathcal{M}(\tilde k) := \ro{Re} \left(\I \langle\partial_{1}\varphi_n(\tilde{k}),(H_{\mathrm{per}}-E_n)(\tilde{k})\partial_{2}\varphi_n(\tilde{k})\rangle_{\mathcal{H}_{\mathrm{f}}}\right)
\]
the Rammal-Wilkinson term. To get a nicer expression we compute the symbol with respect to the $\theta$-quantization. According to Proposition~\ref{alphathetaprop} we have to add
\begin{eqnarray*}\lefteqn{
-\left(\mathcal{A}_{1}(k)\partial_{r_{1}}h^{\rm eff}_{0}(k,r)+ (\mathcal{A}_{2}(k)- \tfrac{\theta k_{1}}{2\pi}  )\partial_{r_{2}}h^{\rm eff}_{0}(k,r)\right)}\\
&=&  - \mathcal{A}_{1}(k)\left(\partial_{r_{1}} E_n(\tilde k)+ \partial_1 \Phi(r) \right)-   \left(\mathcal{A}_{2}(k)- \tfrac{\theta k_{1}}{2\pi} \right)\left(\partial_{2}\Phi(r)-\partial_{1}E_n(\tilde{k})B(r)\right)\,.
\end{eqnarray*}
In summary we have
\begin{eqnarray*}
h^{\rm eff,\theta}_1 (k,r)  &=& - \mathcal{A}_{1}(\tilde{k})\left(  \partial_1 \Phi(r) + \partial_{2}E_n(\tilde{k})B(r) \right) \\
&&\;- \left(\mathcal{A}_{2}(\tilde k)-     \tfrac{\theta k_{1}}{2\pi} \right)\left(\partial_{2}\Phi(r)-\partial_{1}E_n(\tilde{k})B(r)\right)\\
&&\;+B(r) \;\mathcal{M}(\tilde k) \,,
\end{eqnarray*}
where we note that  the combination $\left(\mathcal{A}_{2}(\tilde k)-  \tfrac{\theta k_{1}}{2\pi} \right)$ is a $\Gamma^*$-periodic function.

So in summary we obtain the following corollary.

\begin{cor}\label{h1cor}
Let Assumptions~\ref{A_1: A, Phi, V_Gamma} and \ref{A_2: B0rational} hold with $A^{(1)}=0$ and 
let $E(k)\equiv E_{n}(k) $  be an isolated non-degenerate Bloch band. Then 
there is a $\Gamma^*$-periodic  symbol
$h^{\rm eff, \theta}\in S^{1}(\epsi, \mathcal{L}( \C ))$ such that for the effective Hamiltonian 
$H^{\rm eff}_n := H^{\rm eff}_{I= \{n\}}$ from Theorem~\ref{5.11} it holds that
\begin{equation}\label{claimt1'}
\big\| H^{\mathrm{eff}}_n-{\rm Op}^{\theta}\hspace{-1pt} (h^{\rm eff, \theta})   \big\| _{\mathcal{L}(\mathcal{H}_{\alpha})}= \mathcal{O}(\eps^{\infty})\,.
\end{equation}
The asymptotic expansion of the symbol $h^{\mathrm{eff},\theta}$ can be computed, in principle, to any order in~$\epsi$. Its principal symbol is given by 
\[
h_0(k,r)  =  E (\tilde k )  +\Phi(r )  \,,
\]
and its subprincipal symbol by 
\begin{equation*}
h_1 (k,r)  =  \mathcal{A}(k,r)\cdot  \left(   B(r) \nabla E (\tilde{k})^\perp - \nabla\Phi(r)\right) +B(r) \;\mathcal{M}(\tilde k) \,,
\end{equation*}
where $\tilde k: = k-A(r)$, $\nabla E (\tilde{k})^\perp = ( -\partial_2 E (\tilde k), \partial_1 E (\tilde k))$ and 
\[
\mathcal{M}(  k) = -\ro{Im} \left( \langle\partial_{1}\varphi( {k}),(H_{\mathrm{per}}-E )( {k})\partial_{2}\varphi ( {k})\rangle_{\mathcal{H}_{\mathrm{f}}}\right)\,.
\]
The Berry connection coefficient $\mathcal{A}$  is given by
\[
\mathcal{A}(k,r ) =   \mathcal{A}_1(\tilde k)\,\gamma _1 + \left( \mathcal{A}_{2}(\tilde k)-     \tfrac{\theta  }{2\pi}    \langle k,\gamma_1\rangle\right) \,\gamma _2 \,,
\]
where the components  $\mathcal{A}_j$ are computed from the function $\varphi$ constructed in Proposition~\ref{varphi}
as
\[
\mathcal{A}_j(k) =  - \tfrac{\I}{2\pi} \langle \varphi(k) , \partial_{\kappa_j} \varphi(k)\rangle
:=- \tfrac{\I}{2\pi} \langle \varphi(k) , \gamma_j^*\cdot \nabla \varphi(k)\rangle\,.
\]
\end{cor}

The two terms in the subprincipal symbol have the following physical meaning: Since $\nabla E_n(k)$ is the velocity of a particle with quasi-momentum $k$ in the $n$th band, the term in brackets is the Lorentz force on the particle. Since
the $\theta$-quantization takes into account the integrated curvature of the Berry connection of $2\pi \theta$ per lattice cell of $\Gamma^*$, the curvature form of the effective Berry connection coefficient $\mathcal{A}$ integrates to zero.
 The second term in $h_1$ is  a   correction to the energy known as Rammal-Wilkinson term. For the case $\theta=0$ we recover the first order correction to Peierls substitution established in \cite{PST03b}.

\section{Weyl quantization on the Bloch bundle}\label{quantizations}

In this section we construct quantization  schemes that map suitable symbols to pseudo\-differential ope\-rators that act on sections of possibly non-trivial bundles.
Our construction is related to and motivated by similar constructions in the literature  \cite{Pfl98a,Pfl98b,Saf98,Sha05a,Sha05b,Han10}.
As opposed to the   case of functions on $\R^n$, the relation between a pseudo\-differential operator acting on sections of a vector bundle  and its symbol becomes more subtle. If one defines a corresponding
pseudo\-differential calculus in local coordinates, like this is for example done in \cite{H\"or85}, one can associate a symbol to an operator which is unique only up to an error of order $\eps$.
To define a full symbol, one has to take into account the geometry of the vector bundle. This means that instead of local coordinates, one must use a connection on the vector bundle and a 
connection on the base space. This idea goes back to Widom \cite{Wid78,Wid80}, who was the first to develop a complete isomorphism between such pseudo\-differential operators and their symbols.
However, while he  showed how to recover the full symbol from a pseudo\-differential operator and proved that this map is bijective,  he did not provide an explicit integral formula for the quantization map. His work was developed further by Pflaum \cite{Pfl98b} and Safarov \cite{Saf98}.
In \cite{Pfl98b}, the author   constructs a quantization map which maps symbols that are sections of endomorphism bundles to 
ope\-rators between the sections of the corresponding bundles.
In his quantization formulas he uses a cutoff function so that he can use the exponential map corresponding to a given connection
on the manifold that may not be defined globally. A geometric symbol calculus for pseudo\-differential ope\-rators between sections of vector bundles can also be found in \cite{Sha05a,Sha05b},
where the author moreover introduces the notion of a geometric symbol
in comparison to a coordinatewise symbol. A semiclassical variant of this calculus can be found in \cite{Han10}. When we compute the  symbol $f$   so that
${\rm Op}^{\tau} \hspace{-1pt}(\mathfrak{f})={\rm Op}^{\rm B} \hspace{-1pt}(f )+\Or(\epsi^\infty)$, one could say, using the language of \cite{Sha05a,Sha05b}, that $f $ is the geometric symbol with respect to the Berry connection
of the ope\-rator ${\rm Op}^{\tau} \hspace{-1pt}(\mathfrak{f})$.\\
While in \cite{Saf98} and \cite{Pfl98a}  the authors provide formulas for the Weyl quantization, this is done only for pseudo\-differential ope\-rators
on manifolds and not for ope\-rators between sections of vector bundles. Moreover, the authors   consider only  H\"ormander symbol classes, see \cite{H\"or85}.
In the following we define semiclassical Weyl calculi for more general symbol classes and include the case  of bundles  
with an infinite dimensional Hilbert space as the typical fiber. In addition we   prove a Calderon-Vaillancourt type theorem  establishing $L^2$-boundedness and provide explicit formulas relating the different symbols of an operator corresponding to different quantization maps. However, our constructions are specific to bundles over the torus. Requiring periodicity conditions for symbols and functions allows to project 
the calculus from the cover     $\mathbb{R}^{2}  $ to the quotient $\R^2/\Z^2$,
  an   approach  already  used   in \cite{GN98} and \cite{PST03b,Te03}.
A similar approach was also applied  in \cite{AOS94}, where the authors   consider the Bochner Laplacian  acting  on sections of a line bundle with connection  over the torus. 
 In our calculus the Bochner Laplacian $- \Delta_k$ corresponding to a connection is obtained by quantization of the symbol $f(k,r)=r^{2}$ for $\epsi=1$ using the same connection.

\subsection{The Berry quantization}
The basic idea of the ``Berry quantization'' is to map multiplication by $r$ to the covariant derivative $\I\varepsilon \nabla^{\mathrm{B}}_k$. In contrast to the $\tau$-quantization, where $r$ is mapped to $\I\varepsilon \nabla_k$,
this has two advantages. Since  $\I\varepsilon \nabla^{\mathrm{B}}_k$ is a connection on the Bloch bundle, it leaves invariant its space of  sections.
As a consequence, $f(k, \I\varepsilon \nabla^{\mathrm{B}}_k)$ commutes with $P_I$ if and only if $f(k,r)$ commutes with $P_I(k)$ for all $(k,r)\in M^*\times \R^2$.
 Moreover,  the connection $\nabla^{\mathrm{B}}_k$ restricted to sections
of the Bloch bundle is unitarily equivalent to
the connection $\nabla^{\alpha}_k$ on the bundle $\bu_{\alpha}$ via the unitary map $U_\alpha$.  

As in \cite{PST03b,Te03},
a symbol $f_{\eps}\in S^{w}(\eps,\mathcal{L}(\Hi_{\ro{f}}))$ is called $\tau$-equivariant (more precisely $(\tau_{1},\tau_{2})$-equivariant)
if
$$f_{\eps}(q-\gamma,p)=\tau_{2}(\gamma)f_{\eps}(q,p)\tau_{1}(\gamma)^{-1} \quad \text{for all}\quad \gamma \in \Gamma.$$
The spaces of $\tau$-equivariant symbols are denoted by $S^{w}_{\tau}(\eps,\mathcal{L}(\Hi_{\ro{f}}))$.

Using     the parallel transport $t^{\mathrm{B}}(x,y)$ with respect to the Berry connection introduced in Lemma~\ref{lemma tau equiv varphi}, 
we define the Berry quantization $  {\rm Op}^{\rm B}_\chi\hspace{-1pt} (f)\in \mathcal{L}(\Hi_\tau)$ for $\tau$-equivariant symbols $f\in S^{1}_\tau( \mathcal{L}(\mathcal{H}_{\mathrm{f}}))$ as 
\begin{eqnarray}
\lefteqn{ ({\rm Op}^{\rm B}_\chi\hspace{-1pt} (f)\psi)(k)=}\label{berryQdef}\\ &=& \tfrac{1}{(2\pi \eps)^{2}}\int _{\R^{2}} \left(\int _{\R^{2}} \E^{\tfrac{\I(k-y)r}{\varepsilon}}\chi(k-y)\,t^{\mathrm{B}}\hspace{-1mm}\left(k,\tfrac{k+y}{2}\right)\,f\hspace{-1mm}\left(\tfrac{k+y}{2},r\right)\,t^{\mathrm{B}}\hspace{-1mm}\left(\tfrac{k+y}{2},y\right)\psi(y)\,\mathrm{d}y\right) \mathrm{d}r\,.
\nonumber
\end{eqnarray}
Here, in contrast to the usual Weyl quantization rule,   we  take into account that $\psi$ is a section of a vector bundle with connection $\nabla^{\rm B}$ and the symbol $f(\cdot, r)$ is really a section of its endomorphism bundle. 
So for $f(\frac{k+y}{2},r)$ to act on $\psi(y)$ we first need to map $\psi(y)$ into the correct fiber of the bundle, which is done by the parallel transport $t^{\mathrm{B}}\hspace{-1mm}\left(\tfrac{k+y}{2},y\right)$. However, since the derivatives of $t^{\mathrm{B}}(x,y)$ are not uniformly bounded, we   introduce a cutoff function $\chi$ in the definition. The choice of this cutoff function  has only an effect of order $\Or(\epsi^\infty)$ on the operator, but it simplifies the following analysis considerably. 

\begin{defi}
A function $\chi\in C^{\infty}(\R^{2})$ is called a smooth cutoff function if $\mathrm{supp}\chi$ is compact, $\chi\equiv 1$ in a neighborhood of $0$, and $0\leq \chi \leq 1$.
\end{defi}

Since we need ${\rm Op}^{\rm B}_\chi\hspace{-1pt} (f)$ only for 
$f\in S^{1}_\tau( \mathcal{L}(\mathcal{H}_{\mathrm{f}}))$
as an operator on $\Hi_\tau$, we do not follow the usual routine and show that it is well defined on distributions for general symbol classes. We also do not develop a full Moyal calculus for products of such pseudodifferential operators, although this could be done easily with the tools we provide.

For all steps the following simple lemma will be crucial. It states that the cutoff function in the definition of ${\rm Op}^{\rm B}_\chi\hspace{-1pt} (f)$ ensures that all derivatives of the parallel transport in the integral remain bounded uniformly.

\begin{lem}\label{berrybound}There are constants $c_\alpha$ such that 
\[
\| \partial^\alpha_x t^{\rm B}(x,y) \| \leq c_\alpha \quad\mbox{for all $x,y\in\R^2$ with $|x-y|<1$}\,.
\]
\end{lem}
\begin{pro}
This follows from the smoothness of $t^{\rm B}$ and its $\tau$-equivariance \eqref{tau-equivarianz tB}.
\end{pro}

Before we prove $\Hi_\tau$ boundedness we first show that ${\rm Op}^{\rm B}_\chi\hspace{-1pt} (f)$ is well defined on smooth functions.

\begin{prop}\label{bqprop}
Let $f \in S^{1}_\tau( \mathcal{L}(\mathcal{H}_{\mathrm{f}}))$ and $\psi \in C^\infty(\R^2,\Hi_{\ro{f}})\cap \Hi_\tau$. Then 
${\rm Op}^{\rm B}_\chi\hspace{-1pt} (f) \psi\in C^\infty(\R^2,\Hi_{\ro{f}})\cap \Hi_\tau$.
\end{prop}

\begin{pro}
First note that because of the cutoff function the $y$-integral in \eqref{berryQdef} extends only over a bounded region. Thus one can use
\[
 \E^{-\tfrac{\I  y \cdot r}{\varepsilon}} =  \left( \frac{1- \epsi^2 \Delta_y}{1+r^2} \right)^N \E^{-\tfrac{\I  y \cdot r}{\varepsilon}}
\]
and integration by parts in order to show $r$-integrability of the inner integral. Therefore
$({\rm Op}^{\rm B}_\chi\hspace{-1pt} (f)\psi)(k)$ is well defined and its smoothness follows immediately, since by dominated convergence we can differentiate under the integral and still get enough decay in $r$ by the above trick.  The $\tau$-equivariance of $({\rm Op}^{\rm B}_\chi\hspace{-1pt} (f)\psi)(k)$ can be checked directly using the 
 $\tau$-equivariance of $\psi$, $t^{\rm B}$ and $f$.
\end{pro}

\begin{prop}\label{C-V}
Let $f \in S^{1}_\tau( \mathcal{L}(\mathcal{H}_{\mathrm{f}}))$. Then
${\rm Op}^{\rm B}_\chi\hspace{-1pt} (f)\in \mathcal{L}(\mathcal{H}_{\tau})$     with 
\[
\|{\rm Op}^{\rm B}_\chi\hspace{-1pt} (f)\|_{ \mathcal{L}(\mathcal{H}_{\tau})}\leq c_\chi \|f\|_{\infty,(4,1)}  \,,
   \]
   where the constant $c_\chi$ depends only on $\chi$ and
   \[
   \|f\|_{\infty,(4,1)}    :=  \sum_{|\beta|\leq 4, |\beta'|\leq 1}\sup_{k\in M^*,r\in\R^2}\| \partial^\beta_k\partial^{\beta'}_r f  (k,r)\|_\infty\,.
   \]

\end{prop}
\begin{pro}
Let $\tilde\chi:\R^2\to [0,1]$ be a cutoff function such that supp$\tilde\chi\subset \{|r|<1\}$ and $\sum_{j\in\Z^2} \tilde\chi_{j} (r) \equiv 1$, where $ \tilde\chi_{j} (r) :=  \tilde\chi  (r-j)$, and
let $f_{j} :=  \tilde\chi_{j} f$.
If we can show that $ {\rm Op}^{\rm B}_\chi\hspace{-1pt} (f_j)\in\mathcal{L}(\Hi_\tau)$ and 
\begin{equation}\label{CVbound0}
\sup_{j\in\Z^2} \sum_{i\in\Z^2} \|  {\rm Op}^{\rm B}_\chi\hspace{-1pt} (f_j)^* {\rm Op}^{\rm B}_\chi\hspace{-1pt} (f_i) \|^{\frac12}_{\mathcal{L}(\Hi_\tau)} \leq M\quad\mbox{ and }\quad \sup_{j\in\Z^2} \sum_{i\in\Z^2} \|  {\rm Op}^{\rm B}_\chi\hspace{-1pt} (f_j)   {\rm Op}^{\rm B}_\chi\hspace{-1pt} (f_i)^* \|^{\frac12}_{\mathcal{L}(\Hi_\tau)} \leq M\,,
\end{equation}
then according to the 
  Cotlar-Stein Lemma, cf.\ Lemma~7.10 in \cite{DS99}, it follows that \linebreak
  $\sum_{j\in \Z^2}    {\rm Op}^{\rm B}_\chi\hspace{-1pt} (f_j)$ converges strongly to a bounded operator $F\in \mathcal{L}(\Hi_\tau)$ with $\|F\|_{\mathcal{L}(\Hi_\tau)}\leq M$. However, the following lemma shows that 
  $F=    {\rm Op}^{\rm B}_\chi\hspace{-1pt} (f)$.
 \begin{lem}
Let   $\psi \in C^\infty(\R^2,C^\infty(\R^2,\Hi_{\ro{f}})\cap \Hi_\tau$. Then there is a constant  $C$   such that 
for all $f \in S^{1}_\tau( \mathcal{L}(\mathcal{H}_{\mathrm{f}}))$ with supp$f\subset \R^2\times \{|r|>R\}$
\[
\left\|{\rm Op}^{\rm B}_\chi\hspace{-1pt} (f) \psi(k)\right\|_{\Hi_{\ro{f}}} \leq \frac{C}{R^2} \|f\|_{\infty,(4,0)} .
\]
 \end{lem} 
  \begin{pro}
  Proceed as in the proof of Proposition~\ref{bqprop} using 
  \[
 \E^{-\tfrac{\I  y \cdot r}{\varepsilon}} =  \frac{\epsi^4}{r^4} \,   \Delta_y  ^2 \,\E^{-\tfrac{\I  y \cdot r}{\varepsilon}}
\]
  instead. 
  \end{pro}
  
  Hence on the dense set  $\psi \in C^\infty(\R^2,\Hi_{\ro{f}})\cap \Hi_\tau$
 the sequence $\sum_{j }    {\rm Op}^{\rm B}_\chi\hspace{-1pt} (f_j)\psi $ converges uniformly and thus also in the norm of $\Hi_\tau$ to 
   ${\rm Op}^{\rm B}_\chi\hspace{-1pt} (f )\psi$.

So we are left to show \eqref{CVbound0}, which follows immediately once we can show 
\begin{equation}\label{CVbound}
\|   {\rm Op}^{\rm B}_\chi\hspace{-1pt} (f_j)^*   {\rm Op}^{\rm B}_\chi\hspace{-1pt} (f_i) \|_{\mathcal{L}(\mathcal{\Hi_\tau})}  \leq C ( |i-j| +1)^{-4}\left\|f_{i}\right\|_{\infty,(4,1)}\left\|f_{j}\right\|_{\infty,(4,1)}
\end{equation}
and the analogous second bound for   all $i,j\in\Z^2$. 
Let $\phi,\psi\in \Hi_\tau$, then
\begin{eqnarray*}\lefteqn{\hspace{-1cm}
\langle \phi, (  {\rm Op}^{\rm B}_\chi\hspace{-1pt} (f_j)^*   {\rm Op}^{\rm B}_\chi\hspace{-1pt} (f_i) \psi\rangle_{\Hi_\tau} =
\tfrac{1}{(2\pi \eps)^{4}}
 \int_{M^*} \hspace{-1mm}\D q\int_{\R^8} \D y \, \D k\,\D r\,  \D r'\, \E^{\tfrac{\I k (r-r')}{\varepsilon}}\E^{\tfrac{\I(qr'-yr)}{\varepsilon}}\chi(q-k)\chi(k-y)}\label{expression1}\\&&
\phi^*(q) \,t^{\mathrm{B}}\hspace{-1mm}\left(k,\tfrac{q+k}{2}\right)\,f^*_j\hspace{-1mm}\left(\tfrac{q+k}{2},r'\right)\,t^{\mathrm{B}}\hspace{-1mm}\left(\tfrac{q+k}{2},k\right)
\,t^{\mathrm{B}}\hspace{-1mm}\left(k,\tfrac{k+y}{2}\right)\,f_i\hspace{-1mm}\left(\tfrac{k+y}{2},r\right)\,t^{\mathrm{B}}\hspace{-1mm}\left(\tfrac{k+y}{2},y\right)\psi(y)\,.\nonumber
\end{eqnarray*}
Because of the cutoff functions the domains of integration for $k$ and $y$ are also restricted to compact convex  sets $M^*\subset M_k\subset M_y$ respectively.

For $|i-j|> 2$, $f_i$ and $f_j$ have disjoint $r$-support and 
\[
 \E^{\tfrac{\I k\cdot (r-r')}{\varepsilon}} =  \left( \frac{-\epsi^2\Delta_k}{|r-r'|^2} \right)^2\E^{\tfrac{\I k \cdot (r-r')}{\varepsilon}}\qquad\mbox{for $r-r'\not= 0$}.
\]
Now we insert this into the above integral, integrate by parts, take the norm into the integral and obtain for $|i-j|>2$
\begin{eqnarray*}\lefteqn{
|\langle \phi,  {\rm Op}^{\rm B}_\chi\hspace{-1pt} (f_j)^*   {\rm Op}^{\rm B}_\chi\hspace{-1pt} (f_i) \psi\rangle| \leq}\\&\leq&
\tfrac{\epsi^{4}}{(2\pi \eps)^{4}}
 \int_{M^*} \hspace{-1mm}\D q\int_{M_k}\D k\int_{M_y} \D y \int_{\R^4} \D r\,  \D r'\, \frac{1}{|r-r'|^{4}} \sum_{\beta_1,\cdots,\beta_8} |\partial_k^{\beta_1}\chi(q-k)| |\partial_k^{\beta_2}\chi(k-y)| \times\label{expression2}\\&&\times
\|\phi^*(q)\| \|\partial_k^{\beta_3}t^{\mathrm{B}}\hspace{-1mm}\left(k,\tfrac{q+k}{2}\right)\| \|\partial_k^{\beta_4}f^*_j\hspace{-1mm}\left(\tfrac{q+k}{2},r'\right)\| \|\partial_k^{\beta_5}t^{\mathrm{B}}\hspace{-1mm}\left(\tfrac{q+k}{2},k\right)\|\|\partial_k^{\beta_6}
t^{\mathrm{B}}\hspace{-1mm}\left(k,\tfrac{k+y}{2}\right)\|\times \\&&\times \|\partial_k^{\beta_7}f_i\hspace{-1mm}\left(\tfrac{k+y}{2},r\right)\|\|\partial_k^{\beta_8}t^{\mathrm{B}}\hspace{-1mm}\left(\tfrac{k+y}{2},y\right)\|\|\psi(y)\|\nonumber\\
&\leq& c  \| f_j\|_{\infty,4}\,\| f_i\|_{\infty,4} \sum_{\beta_1, \beta_2}\int_{M^*} \hspace{-1mm}\D q\int_{M_k}\D k\int_{M_y} \D y \int_{{\rm supp}\tilde\chi_i} \D r \int_{{\rm supp}\tilde\chi_j}  \D r'\, \\&&  \qquad\frac{\|\phi(q)\|\|\psi(y)\|}{|r-r'|^{4}} |\partial_k^{\beta_1}\chi(q-k)| |\partial_k^{\beta_2}\chi(k-y)|\,.
\end{eqnarray*}
Here the sum $\sum_{\beta_1,\cdots,\beta_8}$ runs over a finite number of multi-indices 
and we used Lemma~\ref{berrybound}. Moreover we have that because of the $\tau$-equivariance
\[
 \| f_j\|_{\infty,4} := \sum_{|\beta|\leq 4} \sup_{k\in M_y,r\in\R^2}\| \partial^\beta_k f_j (k,r)\|
= \sum_{|\beta|\leq 4} \sup_{k\in M^*,r\in\R^2}\| \partial^\beta_k f  (k,r)\|\,.
\]
For the remaining integral we get 
\begin{eqnarray*}\lefteqn{  \hspace{-0cm}\int_{M^*} \hspace{-1mm}\D q\int_{M_k}\D k\int_{M_y} \D y \int_{{\rm supp}\tilde\chi_i} \D r \int_{{\rm supp}\tilde\chi_j}  \D r'\,   \frac{\|\phi(q)\|\|\psi(y)\|}{|r-r'|^{4}} |\partial_k^{\beta_1}\chi(q-k)| |\partial_k^{\beta_2}\chi(k-y)|}\\&\leq&
 \frac{c_2}{(|i-j|-2)^{4}}\int_{M_k}\D k\, (\|\phi_{M^*}\|*\partial_k^{\beta_1}\chi)(k)
 (\|\psi_{M_y}\|*\partial_k^{\beta_2}\chi)(k)\\
 &\leq&  \frac{c_2}{(|i-j|-2)^{4}} \|\phi_{M^*}\|_2\|\psi_{M_y}\|_2
 \|\partial_k^{\beta_1}\chi\|_1 \|\partial_k^{\beta_2}\chi\|_1
 \leq \frac{c_3}{(|i-j|-2)^{4}} \|\phi\|_{\Hi_\tau}\|\psi\|_{\Hi_\tau}
 \,,
\end{eqnarray*}
where we used Cauchy-Schwarz and Young inequalities in the next to last step. 
Here $\phi_{M^*}(q):= \phi(q) {\bf 1}_{M^* }(q)$ and $\psi_{M_y}(q):= \psi(q) {\bf 1}_{M_y }(q)$.

In order to obtain a bound uniform in $\epsi$
 on $\|   {\rm Op}^{\rm B}_\chi\hspace{-1pt} (f_j)^*    {\rm Op}^{\rm B}_\chi\hspace{-1pt} (f_i) \|_{\mathcal{\Hi_\tau}}$ for all $i,j$ directly, observe that one can get the   factor $\frac{\epsi^4}{|r-r'|^2|k-y||q-k|}$ from   appropriate integrations by parts also  in $r$ and $r'$ using
\[
 \E^{\tfrac{\I  (k-y)\cdot r}{\varepsilon}} =  \left( \frac{-\I \epsi(k-y)\cdot \nabla_r} {|k-y|^2} \right)  \E^{\tfrac{\I  (k-y)\cdot r}{\varepsilon}}\,.
\]
The remaining expression can be bounded as before noting that $\frac{1}{|r-r'|^2}$ is integrable on $\R^4$ and that $\partial_k^\beta\chi(k)/|k|$ is integrable on $\R^2$.
  In summary we can conclude \eqref{CVbound}, which finishes the proof.
\end{pro}

Next we check that the choice of the cutoff function has only an effect of order $\Or(\epsi^\infty)$.

\begin{prop}\label{berry u.a. von cutoff}
Let $f \in S^{1}_\tau( \mathcal{L}(\mathcal{H}_{\mathrm{f}}))$ and let  $\chi_1$ and $\chi_2$ 
be two cutoff functions. Then
$$ \left\|  {\rm Op}^{\rm B}_{\chi_1}\hspace{-1pt} (f )-   {\rm Op}^{\rm B}_{\chi_2}\hspace{-1pt} (f )\right\| = \mathcal{O}(\varepsilon^{\infty}).$$
\end{prop}

\begin{pro}  Let $\bar \chi:= \chi_1-\chi_2$, then $0<c\leq |k|\leq C<\infty$ for all $k\in {\rm supp}\chi$.
We control the norm of $   {\rm Op}^{\rm B}_{\bar \chi }\hspace{-1pt} (f )=  {\rm Op}^{\rm B}_{\chi_1}\hspace{-1pt} (f )-  {\rm Op}^{\rm B}_{\chi_2}\hspace{-1pt} (f )$ as in the previous proof.  So we have to estimate the integrals
\begin{eqnarray*}\lefteqn{\hspace{-1cm}
\langle \phi,  {\rm Op}^{\rm B}_{\bar \chi }\hspace{-1pt} (f_j )^*  {\rm Op}^{\rm B}_{\bar \chi }\hspace{-1pt} (f_i ) \psi\rangle_{\Hi_\tau} =
\tfrac{1}{(2\pi \eps)^{4}}
 \int_{M^*} \hspace{-1mm}\D q\int_{\R^8} \D y \, \D k\,\D r\,  \D r'\, \E^{\tfrac{\I (k-y)\cdot r}{\varepsilon}}\E^{\tfrac{\I(q-k)\cdot r'}{\varepsilon}}\bar\chi(q-k)\bar\chi(k-y)}\label{expression3}\\&&
\phi^*(q) \,t^{\mathrm{B}}\hspace{-1mm}\left(k,\tfrac{q+k}{2}\right)\,f^*_j\hspace{-1mm}\left(\tfrac{q+k}{2},r'\right)\,t^{\mathrm{B}}\hspace{-1mm}\left(\tfrac{q+k}{2},k\right)
\,t^{\mathrm{B}}\hspace{-1mm}\left(k,\tfrac{k+y}{2}\right)\,f_i\hspace{-1mm}\left(\tfrac{k+y}{2},r\right)\,t^{\mathrm{B}}\hspace{-1mm}\left(\tfrac{k+y}{2},y\right)\psi(y)\,.\nonumber
\end{eqnarray*}
Using
\[
 \E^{\tfrac{\I  (k-y)\cdot r}{\varepsilon}} =  \left( \frac{-\epsi^2\Delta_r}{|k-y|^2} \right)^N \E^{\tfrac{\I  (k-y)\cdot r}{\varepsilon}} \qquad\mbox{for $k-y\not= 0$}\,,
\]
we can get any power of $\epsi^2$ by integration by parts and estimating the remaining expression as in the previous proof.
\end{pro}

In the following we drop the subscript $\chi$ in $ {\rm Op}^{\rm B}_{  \chi }\hspace{-1pt} (f )$ in the notation, whenever the statement is not affected by a change of order $\epsi^\infty$. Also note that 
$ {\rm Op}^{\tau} \hspace{-1pt} (f )- {\rm Op}^{\tau}_{  \chi }\hspace{-1pt} (f )= \Or(\epsi^\infty)$ for any cutoff function $\chi$.

Next we relate the $\tau$- and the Berry quantization by using a Taylor expansion of the parallel transport.

\begin{lem}\label{tBexp} For $\delta\in\R^2$ with $|\delta|<\delta_0$ small enough, the parallel transport from $z$ to $z+\delta$  has a uniformly and absolutely convergent expansion 
\[
t^{\rm B}(z+\delta,z) =
 \sum_{n=0}^\infty    t_{n}^{ i_1,\ldots,i_n} (z)\delta_{i_1}\cdots \delta_{i_n} :=  \sum_{n=0}^\infty    \sum_{(i_1,\ldots,i_n)\in \{1,2\}^n}t_{n}^{ i_1,\ldots,i_n} (z)\delta_{i_1}\cdots \delta_{i_n} \,,
\]
where   the coefficients $t_{n}^{ i_1,\ldots,i_n}: \R^2 \to \mathcal{L}(\Hi_{\rm f})$ are  
real-analytic and $\tau$-equivariant.  
The first terms are explicitly
\[
t_0 = {\bf 1}_{\Hi_{\rm f}}\quad\mbox{ and }\quad t_1(z) = M(z) := [\nabla P_I(z),P_I(z)]\,.
\]
\end{lem}
\begin{pro} Note that $t^{\rm B}(z+\delta,z) = t(1)$ where $t(s)$ is the solution of 
\[
\frac{\D}{\D s} t(s) = [ \delta\cdot\nabla P_I(z+s\delta), P_I(z+s\delta)] \,t(s) =: \delta\cdot M(z+s\delta) \,t(s)\quad\mbox{with } t(0)= {\bf 1}\,.
\]
Since $\delta\cdot M: \R^2\to \mathcal{L}(\Hi_{\rm f})$ is smooth and uniformly bounded, the solution of this linear ODE is given by the uniformly convergent Dyson series,
\begin{eqnarray*}\lefteqn{
t^{\rm B}(z+\delta,z) -{\bf 1}  = \sum_{n=1}^\infty   \int_0^1\int_0^{t_1}\cdots\int_0^{t_{n-1}} \delta\cdot M(z+t_1\delta)\cdots \delta\cdot M(z+t_n\delta)\D t_n\cdots\D t_1}\\
&=&\sum_{n=1}^\infty   \int_0^1\int_0^{t_1}\hspace{-3mm}\cdots\hspace{-1mm}\int_0^{t_{n-1}} \sum_{m_1=0}^\infty\hspace{-2mm} \cdots\hspace{-1mm} \sum_{m_n=0}^\infty \frac{t_1^{m_1}(\delta\cdot \nabla)^{m_1}\delta\cdot M(z)}{m_1!}\cdots \frac{t_n^{m_n}(\delta\cdot \nabla)^{m_n}\delta\cdot M(z)}{m_n!}\D t_n\cdots\D t_1\,,
\end{eqnarray*}
where in the second equality we inserted the uniformly convergent power series for the real-analytic function $\delta\cdot M$
\[
\delta\cdot M(z+t\delta) = \sum_{m=0}^\infty \frac{t^{m}(\delta\cdot \nabla)^{m}\,\delta\cdot M(z)}{m!}\,.
\]
\end{pro}

\begin{thm}\label{tauberrythm}
Let $f \in S_{\tau}^{1}( \mathcal{L}(\mathcal{H}_{\mathrm{f}}))$ 
and define for $n\in\N_0$
\[
\mathfrak{f}_n (k,r) := \sum_{\mbox{\tiny$\begin{array}{c} a,b\in\N_0 \\ a+b = n\end{array}$}} \frac{(-1)^a}{(2\I)^n} t_{a}^{ i_1,\ldots,i_a} (k) \;
( \partial_{r_{i_1}}\cdots\partial_{r_{i_a}}\partial_{r_{j_1}}\cdots \partial_{r_{j_b}} f)(k,r)\;
(t_b^{j_1,\ldots,j_b}(k))^*\,.
\]
Then $\mathfrak{f}_n\in  S_{\tau}^{1}( \mathcal{L}(\mathcal{H}_{\mathrm{f}}))$ and 
\begin{equation}\label{compconv}
\Big\|  \sum_{n=0}^N \epsi^n \,{\rm Op}^{\tau} \hspace{-1pt} (\mathfrak{f}_n ) -{\rm Op}^{\rm B} \hspace{-1pt}(f)\Big\|_{\mathcal{L}(\Hi_\tau)} =\Or(\epsi^{N+1})\,.
\end{equation}
The first terms are explicitly $\mathfrak{f}_0(k,r) = f(k,r)$ and
\[
  \mathfrak{f}_1(k,r) =\tfrac{\I}{2} 
\left( \nabla_r f(k,r) \cdot M(k) + M(k) \cdot  \nabla_r f(k,r)
\right)\,,
\]
where $M(k) = [\nabla P_I(k) ,P_I(k)] $.
Moreover, if $f$ has  compact $r$-support, then 
\[
\lim_{N\to \infty} \sum_{n=0}^N \epsi^n \,{\rm Op}^{\tau}_\chi \hspace{-1pt} (\mathfrak{f}_n ) ={\rm Op}^{\rm B}_\chi \hspace{-1pt}(f)\]
strongly in $\Hi_\tau$.

\end{thm}
\begin{pro}
The idea is to insert the Taylor expansion of $t^{\rm B}$ from Lemma~\ref{tBexp} into the integral in Definition~(\ref{berryQdef}). To this end first note that with $\delta:= (k-y)/2$ we have that
\[
t^{\rm B}(k,\tfrac{k+y}{2}) = t^{\rm B}(\tfrac{k+y}{2}+\delta,\tfrac{k+y}{2})\quad\mbox{and}\quad 
t^{\rm B}(\tfrac{k+y}{2},y) = t^{\rm B}(\tfrac{k+y}{2}-\delta,\tfrac{k+y}{2})^*\,.
\]
Assume that $f$ has compact $r$-support for the moment. Then for $\psi\in \Hi_\tau$ we  get
\begin{eqnarray*} 
({\rm Op}^{\rm B}_\chi \hspace{-1pt}(f)\psi)(k)& =& 
\tfrac{1}{(2\pi \eps)^{2}}\int _{\R^{4}}  \mathrm{d}y  \mathrm{d}r \,\E^{\tfrac{\I2\delta\cdot r}{\varepsilon}}\chi(k-y)\, \sum_{a=0}^\infty t_a^{i_1,\ldots, i_a}(\tfrac{k+y}{2}) \delta_{i_1}\cdots\delta_{i_a} \,f\hspace{-1mm}\left(\tfrac{k+y}{2},r\right)\times\\&&\qquad\qquad 
 \sum_{b=0}^\infty (-1)^b \;t_b^{j_1,\ldots, j_b}(\tfrac{k+y}{2})^* \delta_{j_1}\cdots\delta_{j_b}
 \psi(y)
\\
&=&
\tfrac{1}{(2\pi \eps)^{2}}\sum_{a,b=0}^\infty (-1)^b\int _{\R^{4}}  \mathrm{d}y  \mathrm{d}r \,(\tfrac{\epsi}{2\I})^{a+b}\left(  \partial_{r_{i_1}} \cdots \partial_{r_{i_a}} \partial_{r_{j_1}} \cdots 
\partial_{r_{j_b}}
\E^{\tfrac{\I2\delta\cdot r}{\varepsilon}}\right)\times \\&&  \qquad\qquad  \chi(k-y)\,  t_a^{i_1,\ldots, i_a}(\tfrac{k+y}{2})   \,f\hspace{-1mm}\left(\tfrac{k+y}{2},r\right)\, 
 t_b^{j_1,\ldots, j_b}(\tfrac{k+y}{2})^* 
 \psi(y)
 \\
&=&
\tfrac{1}{(2\pi \eps)^{2}}\sum_{a,b=0}^\infty (-1)^a(\tfrac{\epsi}{2\I})^{a+b}\int _{\R^{4}}  \mathrm{d}y  \mathrm{d}r \, \E^{\tfrac{\I(k-y)\cdot r}{\varepsilon}} \chi(k-y)\,  t_a^{i_1,\ldots, i_a}(\tfrac{k+y}{2})  \times\\&&\qquad\qquad 
\left(  \partial_{r_{i_1}} \cdots \partial_{r_{i_a}} \partial_{r_{j_1}} \cdots 
\partial_{r_{j_b}} f\right)\hspace{-1mm}\left(\tfrac{k+y}{2},r\right)\, 
 t_b^{j_1,\ldots, j_b}(\tfrac{k+y}{2})^* 
 \psi(y)\\
 &=& \sum_{n=0}^\infty \epsi^n \,\left( {\rm Op}^{\tau}_\chi \hspace{-1pt} (\mathfrak{f}_n ) \psi\right)(k)\,.
\end{eqnarray*}
Here we used that all sums and integrals converge absolutely and uniformly, so interchanging 
sums and integrals is no problem. Moreover, by the fact that ${\rm Op}^{\rm B}_\chi \hspace{-1pt}(f)\psi$ is a uniformly bounded and $\tau$-equivariant function, the pointwise convergence implies also the strong convergence in $\Hi_\tau$.

In order to estimate 
$\Delta_N\psi:=  \left( \sum_{n=0}^{N-1} \epsi^n \,{\rm Op}^{\tau}_\chi \hspace{-1pt} (\mathfrak{f}_n ) - {\rm Op}^{\rm B}_\chi \hspace{-1pt}(f)\right) \psi$ in $\Hi_\tau$, we estimate as in the previous proofs $|\langle \phi, \Delta_N \psi\rangle |$.
Write for the remainder in the Taylor expansion
\begin{eqnarray*}
t^{\rm B}(z+\delta,z) &=&
 \sum_{a=0}^{N-1}    t_{a}^{ i_1,\ldots,i_a} (z)\delta_{i_1}\cdots \delta_{i_a} + 
 \frac{(\partial_{i_1} \cdots \partial_{i_{N}} t^{\rm B} )(z+ \xi(\delta)\delta,z) }{N!}\delta_{i_1}\cdots \delta_{i_{N}} \\
 &=: & \sum_{a=0}^{N-1}    t_{a}^{ i_1,\ldots,i_a} (z)\delta_{i_1}\cdots \delta_{i_a} +  R_{N}^{i_1,\ldots, i_{N}} (z,\delta) \delta_{i_1}\cdots \delta_{i_{N}} \,,
\end{eqnarray*}
then one term appearing in the estimate of $|\langle \phi, \Delta_N \psi\rangle |$ is
\begin{eqnarray*} 
\lefteqn{ 
\tfrac{1}{(2\pi \eps)^{2}}\int_{M^*}\D k \int _{\R^{4}}  \mathrm{d}y  \mathrm{d}r \,\E^{\tfrac{\I2\delta\cdot r}{\varepsilon}}\chi(k-y)\, \phi^*(k) 
 R_{N}^{i_1,\ldots, i_{N}}  
(\tfrac{k+y}{2},\delta) \delta_{i_1}\cdots\delta_{i_N} \,f\hspace{-1mm}\left(\tfrac{k+y}{2},r\right)
 \psi(y)}
\\
&=&\tfrac{1}{(2\pi \eps)^{2}} \left( \tfrac{-\epsi}{2\I}\right)^{N}\int_{M^*}\D k \int _{\R^{4}}  \mathrm{d}y  \mathrm{d}r \,\E^{\tfrac{\I(k-y) \cdot r}{\varepsilon}}\chi(k-y)\, \phi^*(k) 
 R_{N}^{i_1,\ldots, i_{N}}  
(\tfrac{k+y}{2},\delta) \\&& \qquad\qquad \left(  \partial_{r_{i_1}} \cdots \partial_{r_{i_N}}   f\right)\hspace{-1mm}\left(\tfrac{k+y}{2},r\right)    \psi(y)\,.
\end{eqnarray*}
Such an expression can be bounded by a constant  times $\epsi^{N}  \|\phi\|\|\psi\|$ by obtaining an integrable factor $\frac{\epsi^2}{|r||k-y|}$ through additional integration by parts as in the proof of Proposition~\ref{C-V}. All other terms can be treated similarly and we have shown \eqref{compconv} for $f$ with compact $r$-support.

For the general statement we use again the Cotlar-Stein lemma on the family of almost orthogonal operators $\Delta_{N,i} :=\sum_{n=0}^{N-1} \epsi^n \,{\rm Op}^{\tau}_\chi \hspace{-1pt} (\mathfrak{f}_{n,i} ) - {\rm Op}^{\rm B}_\chi \hspace{-1pt}(f_i)$. While this is very lengthy to write down, the estimates are completely analogous to those of Proposition~\ref{C-V} using integration by parts as before.
\end{pro}

Of course we can reverse the rolls of the two quantizations and obtain the reverse statement.

\begin{cor}\label{tauberry}
Let $\mathfrak{f} \in S_{\tau}^{1}( \mathcal{L}(\mathcal{H}_{\mathrm{f}}))$ 
and define 
\[
f_n (k,r) := \sum_{a+b = n} \frac{(-1)^a}{(2\I)^n} (t_{a}^{ i_1,\ldots,i_a} (k))^* \;
( \partial_{r_{i_1}}\cdots\partial_{r_{i_a}}\partial_{r_{j_1}}\cdots \partial_{r_{j_b}} \mathfrak{f})(k,r)\;
t_b^{j_1,\ldots,j_b}(k) \quad\mbox{for $n\in\N_0$}\,.
\]
Then $f_n\in  S_{\tau}^{1}(\R^{4},\mathcal{L}(\mathcal{H}_{\mathrm{f}}))$ and
\[
\Big\|  \sum_{n=0}^N \epsi^n \,{\rm Op}^{\rm B} \hspace{-1pt}(f_n) - {\rm Op}^{\tau} \hspace{-1pt}(\mathfrak{f})\Big\|_{\mathcal{L}(\Hi_\tau)} =\Or(\epsi^{N+1})\,.
\]
 The first terms are explicitly $f_0(k,r) =\mathfrak{f}(k,r)$ and
\[
  f_1(k,r) =-\tfrac{\I}{2} 
\left( \nabla_r \mathfrak{f}(k,r) \cdot M(k) + M(k) \cdot  \nabla_r \mathfrak{f}(k,r)
\right)\,.
\]

\end{cor}

While we do not use the following proposition explicitly, it sheds some light on the geometric significance of the Berry quantization. It states that ${\rm Op}^{\rm B} \hspace{-1pt} (f)$ commutes with the projection $P_I$ if and only if  the symbol $f(k,r)$ commutes pointwise with $P_I(k)$.

\begin{prop}\label{fPI0=0}
Let $f \in S_{\tau}^1(\mathcal{L}(\mathcal{H}_{\mathrm{f}}))$. Then  
$$[f(k,r),P_I(k)]=0 \quad  \forall (k,r) \in \R^{4} \quad \Leftrightarrow\quad [{\rm Op}^{\rm B} \hspace{-1pt} (f),P_I]  =0 \,. $$
\end{prop}

\begin{pro}  It suffices to consider the commutator on the dense set $C^\infty(\R^2,\Hi_{\ro{f}})\cap \Hi_\tau$, so we can work with the integral definition \eqref{berryQdef} of ${\rm Op}^{\rm B} \hspace{-1pt} (f)$. For $\psi\in C^\infty(\R^2,\Hi_{\ro{f}})\cap \Hi_\tau$ it follows from  \eqref{berryconP} 
that
\begin{eqnarray*}
\lefteqn{ \left(\left[ {\rm Op}^{\rm B}_\chi\hspace{-1pt} (f),P_I\right] \psi\right)(k)=}\\ && \tfrac{1}{(2\pi  )^{2}}\int _{\R^{2}} \left(\int _{\R^{2}} \E^{ \I(k-y)r}\chi(k-y)\,t^{\mathrm{B}}\hspace{-1mm}\left(k,\tfrac{k+y}{2}\right)\,\left[ f\hspace{-1mm}\left(\tfrac{k+y}{2},\epsi r\right), P_I\hspace{-1mm}\left(\tfrac{k+y}{2}\right)\right] \,t^{\mathrm{B}}\hspace{-1mm}\left(\tfrac{k+y}{2},y\right)\psi(y)\,\mathrm{d}y\right) \mathrm{d}r
\nonumber
\end{eqnarray*}
so 
the  implication from left to right is obvious. To prove the reverse implication in detail is somewhat tedious. Since we don't use it, we only sketch the argument. Assume that $[f(k,r),P_I(k)]= O(k,r)\not=0$. Then $O\in S_{\tau}^1(\R^{4},\mathcal{L}(\mathcal{H}_{\mathrm{f}}))$ and one can  show that $\| {\rm Op}^{\rm B}_\chi\hspace{-1pt} (O)\| \geq C>0 $ for some $C$ independent of $\epsi$ by looking at the action of $ {\rm Op}^{\rm B}_\chi\hspace{-1pt} (O)$ on suitable coherent states.
This implies even the stronger statement
$$ [{\rm Op}^{\rm B} \hspace{-1pt} (f),P_I]  =o(\epsi )
 \quad \Rightarrow\quad  [f(k,r),P_I(k)]=0 \quad  \forall (k,r) \in \R^{4}\,.$$
 \end{pro}

\subsection{The $\alpha$-quantization and the $\theta$-quantization}

The other two quantizations we use are the $\alpha$-quantization and the effective quantization. The $\alpha$-quantization   with respect to the connection $\nabla^\alpha  = U_\alpha \nabla^{\rm B} U_{\alpha}^*$ is used to map $\alpha$-equivariant  symbols in $C^{\infty}(\R^{4},\mathcal{L}(\C^m))$ to operators in $\mathcal{L}(\Hi_\alpha)$, c.f.\ Definition~\ref{uthetadef}. For $m=1$ it can be replaced by  the effective quantization with respect to 
 the explicit connection $\nabla_k^{\theta} := \nabla_{k}+   \tfrac{\I \theta}{2\pi}  \langle k,\gamma_1 \rangle \,\gamma_2$.

In both cases the construction is exactly the same as the one for the Berry quantization, which is to use the parallel transport of the desired  connection in the
definition of the quantization.  Let 
\[
t^{\alpha}(x,y):\C^m\to \C^m \,, \quad\lambda \mapsto t^{\alpha}(x,y)\lambda
  := U_{\alpha}(x)t^{\rm B}(x,y)U_{\alpha }^*(y)\lambda
 \]
 be the parallel transport along the straight line from $y$ to $x$ with respect to the connection $\nabla^{\alpha}=U_{\alpha}\nabla^{\mathrm{B}}U_{\alpha}^{*}$. Then $\tau$-equivariance of~$t^{\rm B}$  implies $\alpha$-equivariance of~$t^\alpha$, i.e.\ 
 \[
 t^{\alpha}(x-\gamma^*,y-\gamma^*)= \alpha\hspace{-2pt}\left(\tfrac{\langle x,\gamma_2\rangle}{2\pi}\right)^{-n_1} t^{\alpha}(x ,y )\;\alpha\hspace{-2pt}\left(\tfrac{\langle y,\gamma_2\rangle}{2\pi}\right)^{n_1}\,.
 \]
For $m=1$ we   introduce  the effective connection $\nabla^{\theta}_{k}=\nabla _{k} +  \tfrac{\I  \theta}{2\pi} \langle k,\gamma_1\rangle \,\gamma_2$ and the corresponding $\alpha$-equivariant  parallel transport 
\[
t^{\theta}(x,y):\C \to \C  \,, \quad\lambda \mapsto t^{\theta}(x,y)\lambda
 := \E^{\frac{\I\theta}{4\pi}\langle x +y,\gamma_1\rangle  \langle y-x,\gamma_2\rangle}\lambda\,.
\]
We say that a symbol $f \in C^\infty(\R^{4},\mathcal{L}(\C^m))$
is $\alpha$-equivariant, if
\[
f(k-\gamma^*,r) = \alpha(\kappa_2)^{-n_1} f(k,r) \alpha(\kappa_2)^{n_1}\quad \mbox{for all $\gamma^*\in\Gamma^*$, $k,r\in\R^2$\,,}
\]
where we use again the notation $\kappa_j = \frac{\langle k, \gamma_j\rangle}{2\pi}$.
Note that for $m=1$ the $\alpha$-equivariant symbols are just the periodic symbols. However, for $m>1$ the $\kappa_2$-derivatives of an $\alpha$-equivariant symbol are in general unbounded as functions of $\kappa_1$.
Thus we define the space of ``bounded'' symbols  $S_\alpha( \mathcal{L}(\C^m))$ as follows.
\begin{defi} \label{Salpha} Let $S_\alpha( \mathcal{L}(\C^m))$ be
the space of $\alpha$-equivariant functions $f \in C^\infty(\R^{4},\mathcal{L}(\C^m))$ that satisfy
\[
  \sup_{k\in M^*,r\in\R^{2}}    \| (\partial_k^\alpha \partial_r^\beta f)(k,r)\|_{\mathcal{L}(\C^m)}  < \infty   \quad\mbox{ for all $\alpha,\beta \in \N^2_0$.}
\]
As always, $S_\alpha(\mathcal{L}(\C^m))$ is equipped with the corresponding Fr\'echet metric and $S_\alpha(\epsi,\mathcal{L}(\C^m))$ denotes the space of uniformly bounded functions $f:[0,\epsi_0)\to S_\alpha(\mathcal{L}(\C^m))$.
\end{defi}

In complete analogy to the Berry quantization we define for $\alpha$-equivariant symbols 
$f \in S_\alpha(\mathcal{L}(\C^m))$ and $\psi\in \Hi_\alpha$ 
 the $\alpha$-quantization by
$$\left({\rm Op}^{\alpha}_{\chi}(f) \psi\right)(k):= \tfrac{1}{(2\pi \eps)^{2}}\int _{\R^{2}} \left(\int _{\R^{2}} \E^{\tfrac{\I(k-y)r}{\varepsilon}}\chi(k-y)\,t^{\alpha}\hspace{-1mm}\left(k,\tfrac{k+y}{2}\right)\,f\hspace{-1mm}\left(\tfrac{k+y}{2},r\right)\,t^{\alpha}\hspace{-1mm}\left(\tfrac{k+y}{2},y\right)\psi(y)\,\mathrm{d}y\right) \mathrm{d}r \,,
$$
and for $m=1$ 
 the $\theta$-quantization by
\begin{eqnarray*}
\left({\rm Op}^{\theta}_{\chi}(f) \psi\right)(k)&\hspace{-1mm}:=& \hspace{-1mm}\tfrac{1}{(2\pi \eps)^{2}}\int _{\R^{2}} \left(\int _{\R^{2}} \E^{\tfrac{\I(k-y)r}{\varepsilon}}\chi(k-y)\,t^{\theta}\hspace{-1mm}\left(k,\tfrac{k+y}{2}\right)\,f\hspace{-1mm}\left(\tfrac{k+y}{2},r\right)\,t^{\theta}\hspace{-1mm}\left(\tfrac{k+y}{2},y\right)\psi(y)\,\mathrm{d}y\right) \mathrm{d}r \\
&\hspace{-1mm}=&\hspace{-1mm}
\tfrac{1}{(2\pi \eps)^{2}}\int _{\R^{2}} \left(\int _{\R^{2}} \E^{\tfrac{\I(k-y)r}{\varepsilon}}\chi(k-y)\,\E^{\frac{\I\theta}{4\pi}\langle x +y,\gamma_1\rangle  \langle y-x,\gamma_2\rangle} \,f\hspace{-1mm}\left(\tfrac{k+y}{2},r\right) \psi(y)\,\mathrm{d}y\right) \mathrm{d}r\,.
\end{eqnarray*}

Now we can show all results of the previous section in a completely analogous way also for the $\alpha$- and the $\theta$-quantization. 

\begin{prop}\label{C-Valpha}
Let $f \in S_\alpha (\R^{4},\mathcal{L}(\C^m))$. Then
${\rm Op}^{\alpha}_\chi\hspace{-1pt} (f)\in \mathcal{L}(\mathcal{H}_{\alpha})$     with 
\[
\|{\rm Op}^{\alpha}_\chi\hspace{-1pt} (f)\|_{ \mathcal{L}(\mathcal{H}_{\alpha})}\leq c_\chi \|f\|_{\infty,(4,1)}    :=c_\chi \sum_{|\beta|\leq 4, |\beta'|\leq 1}\sup_{k\in M^*,r\in\R^2}\| \partial^\beta_k\partial^{\beta'}_r f  (k,r)\|\,,
   \]
   where the constant $c_\chi$ depends only on $\chi$. For $m=1$ the same bound holds for ${\rm Op}^{\theta}_\chi\hspace{-1pt} (f)$.
\end{prop}

\begin{prop}\label{theta berry}
Let $f \in  S_{\tau}^{1}(\mathcal{L}(\mathcal{H}_{\mathrm{f}}))$ and
\[
f_I(k,r)_{ij} := \langle \varphi_i(k), f(k,r) \varphi_j(k)\rangle\,.
\]
Then $f_I \in S_\alpha(\C^m)$ and
\[
{\rm Op}^{\alpha}_\chi\hspace{-1pt} (f_I) =U_\alpha {\rm Op}^{\rm B}_\chi\hspace{-1pt} (f)\,
U_\alpha^*\,.
\]
\end{prop}
\begin{pro}
It follows directly from the definitions that $f_I \in S_\alpha(\C^m)$. The equality of the operators  can be checked on the dense set $C^\infty(\R^2)\cap \Hi_\alpha$ using their integral definitions  and the fact that, again by definition,
$U_{\alpha}^*(x)t^{\alpha}(x,y) 
   = t^{\rm B}(x,y)U_{\alpha }^*(y) $.
\end{pro}

For the case $m=1$ we can finally replace the $\alpha$- by the $\theta$-quantization if we suitably modify the  symbol. To this end we introduce   the Taylor series of the difference of the parallel transports as
\[
t^{\theta *}(k,k+\delta) t^\alpha(k,k+\delta) =:  \sum_{n=0}^\infty    t_{n}^{ i_1,\ldots,i_n} (k)\delta_{i_1}\cdots \delta_{i_n}\,,
\]
where 
\[
t_0 (k) \equiv  1\quad\mbox{and}\quad t_1 (k) = \I\left(   \mathcal{A}_{1}(k)\gamma_1 +  \left(    \mathcal{A}_{2}(k)-  \tfrac{\theta}{2\pi} \langle k, \gamma_1\rangle  \right)\gamma_2
\right) = : \I \mathcal{A}(k) \,.
\]
The proof of the following proposition is analogous to the proof of Theorem~\ref{tauberrythm}. The expressions simplify a bit because for $m=1$ the symbol and the parallel transport commute.

\begin{prop}\label{alphathetaprop}
Let $f \in S^{1}(\R^{4},\C)$ be a periodic symbol and
and define for $n\in\N_0$
\[
 {f}_n^\theta (k,r) :=  \I^n\, t_{n}^{ i_1,\ldots,i_n} (k) \;
( \partial_{r_{i_1}}\cdots\partial_{r_{i_n}}  f)(k,r)
\,.
\]
Then $ {f}_n^\theta \in  S^{1}(\R^{4},\C)$ is periodic and 
\begin{equation}\label{compconv'}
\Big\|  \sum_{n=0}^N \epsi^n \,{\rm Op}^{\theta} \hspace{-1pt} ( {f}_n^\theta ) -{\rm Op}^{\alpha} \hspace{-1pt}(f)\Big\|_{\mathcal{L}(\Hi_\alpha)} =\Or(\epsi^{N+1})\,.
\end{equation}
The first terms are explicitly $ {f}_0^\theta(k,r) = f(k,r)$ and
\[
  {f}_1^\theta(k,r) =-\mathcal{A} (k) \cdot \nabla_r f(k,r) \,.
\]
 
\end{prop}

\section{Application to the Hofstadter model}

In this section we apply the general theory developed in the previous sections to perturbations of magnetic subbands of the Hofstadter Hamiltonian \cite{Hof76}. The motivation for doing this is twofold. First it shows in the simplest possible example how magnetic Peierls substitution Hamiltonians can be explicitly computed and analyzed. Second, we will find strong support for the conjecture that   Theorem~\ref{5.11} is actually still valid for perturbations by small constant fields~$B$.
Note that the Hofstadter Hamiltonian and related tight binding models served not only as model Hamiltonians for the illustration of general results on perturbed periodic Schr\"odinger operators, but 
also gave rise to considerable mathematical work dedicated  specifically to them, e.g.\ 
\cite{HS89,HS90a,HKS90,BKS91,AJ09}. For a recent overview of the mathematics and the physics literature on the Hofstadter Hamiltonian we refer to \cite{DeN10}.

\begin{figure}
\floatbox[{\capbeside\thisfloatsetup{capbesideposition={right,center},capbesidewidth=6cm}}]{figure}[9.5cm]
{\caption{\small The black and white butterfly \cite{Hof76} showing the spectrum of  $H^{B_0}_{\rm Hof}$  as a function of $B_0$.  For rational values $B_0=2\pi \frac{p}{q}$ the spectrum of $H^{B_0}_{\rm Hof}$ consists of $q$ disjoint intervals if $q$ is odd and of $q-1$ disjoint intervals if $q$ is even. }\label{BW}}
{\includegraphics[width=9.5cm]{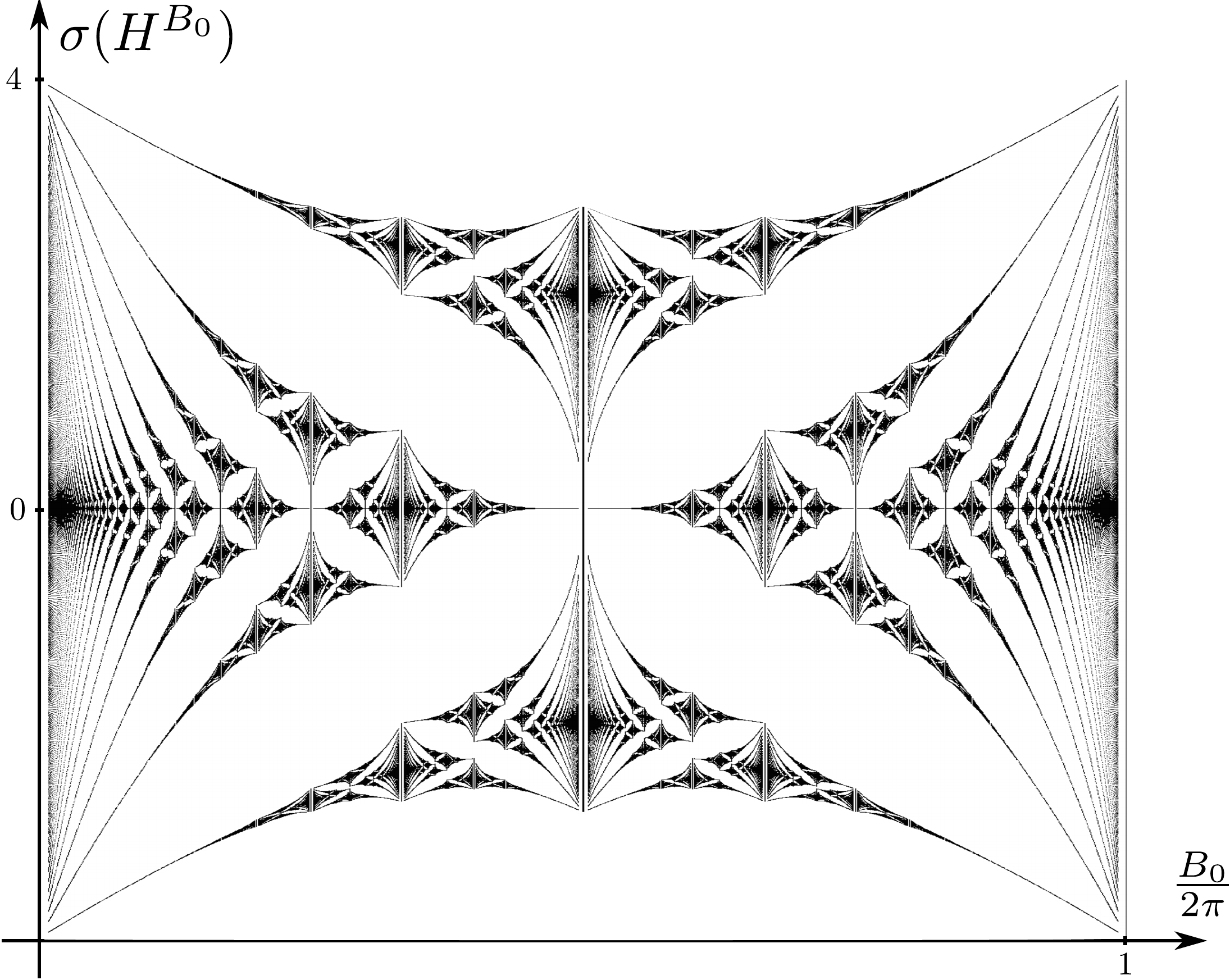}}
\end{figure}

The Hofstadter model is the canonical model for a single non-magnetic Bloch band perturbed by a constant magnetic field $B_0$. It can be seen to arise from 
 the tight-binding formalism in physics or, alternatively, from Peierls substitution for a  non-magnetic Bloch band.
  The Hofstadter Hamiltonian is the discrete magnetic Laplacian on the lattice $\widetilde\Gamma = \Z^2$, 
\[
H^{B_0 }_{\ro{Hof}} = D_1 +  D_1^*  + D_2 + D_2^* \quad\mbox{acting on} \quad \ell^2(\Z^2)\,.
\]
Here 
 $D_1$ and $D_2$ are     the  (dual)  magnetic translations
  \begin{equation*} 
(D_1 \psi)(x):=  \psi(x- e_1)\quad\mbox{and}\quad 
(D_2 \psi)(x):=   \E^{\I B_0\langle x, e_1\rangle  } \psi(x- e_2)\,.
\end{equation*} 
For $B_0 = 2\pi p/q$ we define the corresponding magnetic Bloch-Floquet transformation on the lattice $\Gamma = q\Z\times \Z$ as
\[
\mathcal{U}_{\mathrm{BF}}: \ell^2(\Gamma;\C^q)\to L^2(\T^*_q;\C^q)\,,\quad(\mathcal{U}_{\mathrm{BF}}\psi)(k)_j := \sum_{\gamma \in \Gamma} \E^{\I  \gamma \cdot k}(T_{\gamma}\psi)((j,0))\quad\mbox{for $j=0,\ldots,q-1$}\,,
\]
where we recall that the magnetic translations $T_\gamma$ were defined in \eqref{TDef}.
Note that the fiber space $\Hi_{\rm f}=\C^q$ is now finite dimensional and thus we can drop the additional phase $\E^{-\I k\cdot y}$    in the   definition of $\mathcal{U}_{\mathrm{BF}}$, which appeared in  \eqref{UBF} to make the domain of $H_{\rm per}(k)$ independent of $k$. As a consequence, the range of $\mathcal{U}_{\mathrm{BF}}$ now contains periodic functions on $\T^*_q = [0,\frac{2\pi}{q})\times [0,2\pi)$ and $\tau$-equivariance becomes periodicity.
A straightforward computation shows that the shift operators $D_j$ become matrix-multiplication operators    $\hat D_j := \mathcal{U}_{\mathrm{BF}}\,D_j\,\mathcal{U}_{\mathrm{BF}}^*$,
\[
  \hat D_1 (k)= 
\begin{pmatrix}
0 & 0 & 0& \cdots& \E^{\I q k_1}\\
1&0&0&\cdots&0\\
0&1&0&\cdots &0\\
\vdots&\vdots&\ddots&\ddots&\vdots\\
0&0&\cdots&1&0
\end{pmatrix}\,,\quad
\hat D_2 (k) = 
\E^{\I k_2}
\begin{pmatrix}
1 & 0&0&\cdots&0\\
0& \E^{\I B_0}& 0&\cdots&0\\
0&0&\E^{\I 2B_0}&\ddots&0\\
\vdots&\vdots&\ddots&\ddots&0\\
0&0&\cdots&0&\E^{\I (q-1)B_0}
\end{pmatrix}\,.
\]
For the Hamiltonian one thus finds
\[
\hat H^{B_0}_{\rm Hof}(k)   =\scriptstyle\small
\begin{pmatrix}
2 \cos (k_{2}) &1 & 0&\ldots&\E^{\I qk_{1}}\\[1mm]
1 & 2 \cos (k_{2}-B_0) &1&\ldots&0\\[1mm]
0 &1& 2 \cos (k_{2}-2B_0) &\ldots&0\\
\vdots& \ddots& \ddots&\ddots&0\\[1mm]
0&\ &\ &\ &1\\[1mm]
\E^{-\I qk_{1}}&0&\ldots&1&2 \cos (k_{2}-(q-1)B_0)\\
\end{pmatrix}\,,
\]
which is indeed $2\pi/q$-periodic in $k_1$ and $2\pi$-periodic in $k_2$. The spectrum of $\hat H^{B_0}_{\rm Hof}(k)$ consists of $q$ distinct eigenvalue bands $E_n(k)$, $n=1,\ldots, q$, with periodic spectral projections $P_n(k)$, defining the magnetic Bloch bands and Bloch bundles of the Hofstadter model. The spectrum of $H^{B_0}_{\rm Hof}$ is the union of the ranges of the functions $E_n(k)$ and thus consists of $q$ intervals. As a function of $B_0$ the spectrum is depicted in the famous Hofstadter butterfly, Figure~\ref{BW}. Note that for $B_0\notin 2\pi \Q$ the spectrum of $H^{B_0}_{\rm Hof}$ is a Cantor-type set, i.e.\ a nowhere dense, closed set of Lebesgue measure zero, cf.\ \cite{AJ09}.

\begin{figure}
\floatbox[{\capbeside\thisfloatsetup{capbesideposition={right,center},capbesidewidth=6cm}}]{figure}[9.5cm]
{\caption{\small The colored butterfly for the Hofstadter Hamiltonian $H^{B_0}_{\rm Hof}$, as first plotted in \cite{OA01}.   The colored regions are open components of the resolvent set and the colors encode Chern numbers of overlying Bloch bundles. Physically, the Chern numbers represent the Hall conductivity of a corresponding non-interacting Fermi gas. For fixed $B_0$, i.e.\ in each vertical line, the Chern numbers of the single bands sum up to the total Chern number $\theta=0$, as represented by the white region on bottom of the butterfly.}\label{Colors}}
{\includegraphics[width=9.5cm]{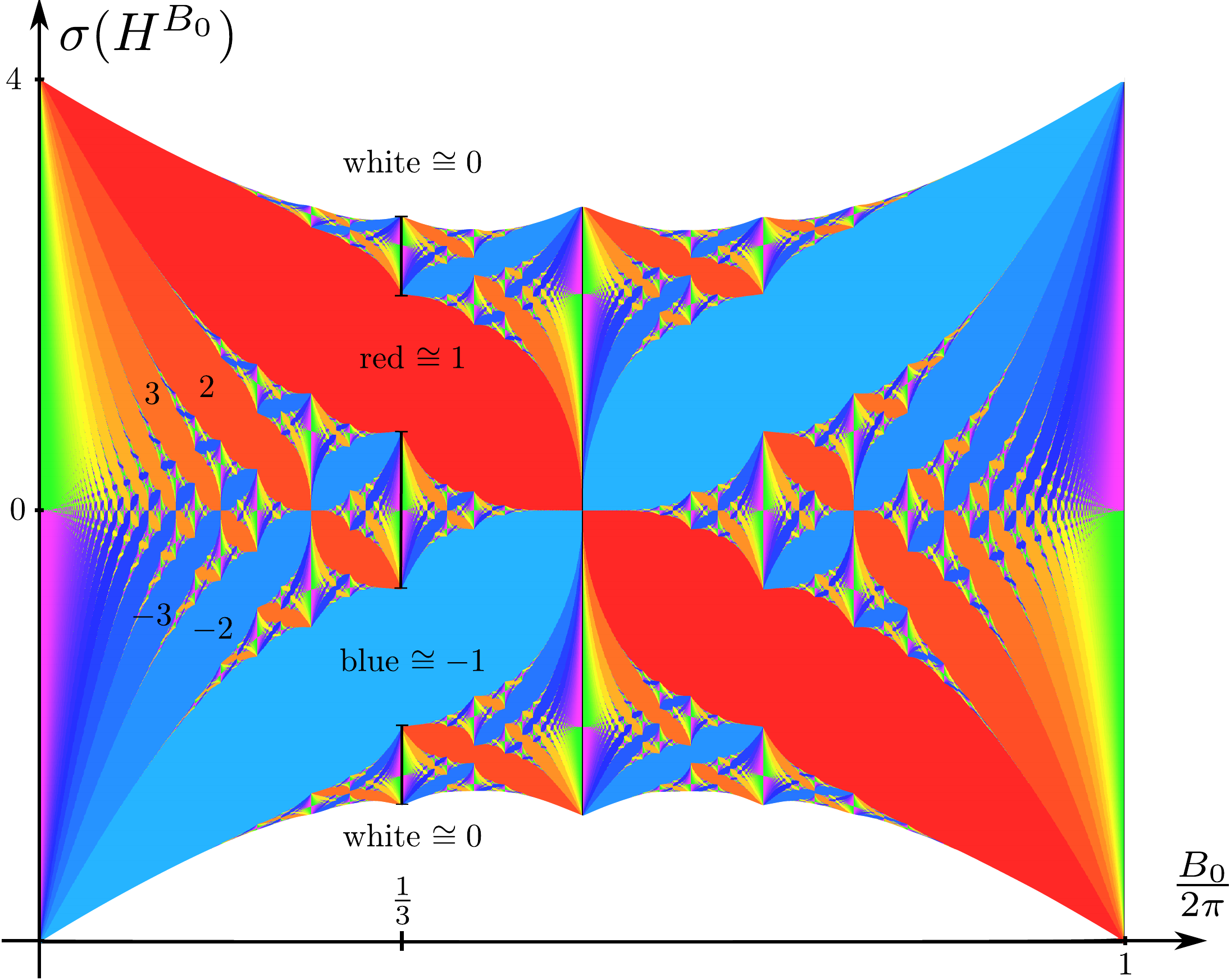}}
\end{figure}

Osadchy and Avron \cite{OA01} produced a colored version of the butterfly    by coloring the gaps in the spectrum according to the sum of the Chern numbers of the  overlying bands, see Figure~\ref{Colors}. E.g.\ for $B_0=2\pi \frac13$, the top and the bottom bands have Chern number $1$ each and the middle band has Chern number $-2$. Thus the gaps are labeled from top to bottom by $0$ (white), $1$ (red), $-1$ (blue), and again $0$ (white).

Now we  apply the machinery developed in the previous sections to determine  Peierls substitution Hamiltonians for magnetic subbands of $H^{B_0}_{\rm Hof}$. Let $B_0 = 2\pi \frac{p}{q}$, then $\hat H^{B_0}_{\rm Hof}(k)$ is a matrix valued function on the torus $\T^*_q = [0,\frac{2\pi}{q})\times [0,2\pi)$, but its eigenvalue bands have period $2\pi/q$ in both directions. Hence  we can take  as a model dispersion relation  
\[
E_q (k) := 2\left( \cos(qk_1) +  \cos(qk_2) \right)= \E^{\I qk_1} +\E^{-\I qk_1}+\E^{\I qk_2}+\E^{-\I qk_2}\,.
\]
This is, up to a constant factor, the leading order part in the Fourier expansion of any Bloch band $E_n(k)$ on $\T^*_q$.
So we pick an isolated simple subband of $\hat H^{B_0}_{\rm Hof}(k)$ with Chern number $\theta\in \Z$ and approximate its dispersion by $E_q(k)$. 
If we now perturb  $B_0$ by an additional ``small'' constant magnetic field $B = {\rm curl}A(x)$ with $A(x) = (0, B x_1)$, the   Peierls substitution Hamiltonian for this subband is given as  the $\theta$-quantization of $E_q(k-A(r))$,
\[
H^B_{\theta,q}  := {\rm Op}^\theta(E_q(k-A (r)))  =  \E^{\I q\mathcal{K}_1 } + \E^{-\I q\mathcal{K}_1 } + \E^{\I q\mathcal{K}_2 } + \E^{-\I q\mathcal{K}_2 }
\]
with 
\[
\mathcal{K}_1  = k_1  
\quad \mbox{ and } \quad 
\mathcal{K}_2 = k_2  -\I B\nabla^\theta_{1} = k_2  -\I B\partial_{k_1}
\]
acting on  
\[
 \mathcal{H}_\theta  =  \left\{ f \in L^{2}_{\mathrm{loc}}(\R ^{2} )\,|\,f(k_1-\tfrac{2\pi}{q},k_2)=\E^{   \I   \theta k_{2} }  f(k_1,k_2)\,,\; f(k_1,k_2-2\pi) =f(k_1,k_2) \right\}\,.
\]
Here $\nabla^\theta_k = (\partial_{k_1}, \partial_{k_2} + \I\frac{q \theta k_1}{2\pi})$ and due to our choice of gauge for the perturbing magnetic field the operator $H^B_{\theta,q} $ depends on $\theta$ only through its domain.
Note  that this gauge is different from the one used in Theorem~\ref{5.11} and we use it to simplify the analysis of the resulting operator  $H^B_{\theta,q} $. However,  since 
   Theorem~\ref{5.11} does not cover the case of a perturbation by a constant magnetic field anyways,  our derivation of $H^B_{\theta,q}$ is merely heuristic for any choice  of gauge.  
 
  To determine the spectrum of  $H^B_{\theta,q}$,  it is sufficient to notice  that it has the structure
\[
  U_1 + U_1^* + U_2 + U_2^*
\]
with unitary operators $U_1 $ and $U_2 $ that satisfy 
\begin{equation}\label{NCT}
U_1U_2= \E^{ \I q^2 B  }U_2U_1 =:  \E^{ \I \alpha  }U_2U_1\,.
\end{equation}
The $C^*$-algebra $\mathcal{N}_\alpha$ generated by   two abstract  elements $\mathfrak{U}_1$ and $\mathfrak{U}_2$ satisfying \eqref{NCT} is called the non-commutative torus. The  mappings 
\[
\pi^B_{\theta,q}: \mathcal{N}_{q^2B}\to \mathcal{L}(\Hi_\theta)\,,\; \mathfrak{U}_j \mapsto \E^{\I q \mathcal{K}_j }
\]
  thus define a $*$-representation  of $\mathcal{N}_{q^2B}$ into the bounded operators on   $\Hi_\theta$. Accordingly,  each   operator  $H^B_{\theta,q}$ is a representation  of the   abstract element $\mathfrak{H}^\alpha=  \mathfrak{U}_1 + \mathfrak{U}_1^* + \mathfrak{U}_2 + \mathfrak{U}_2^*$ of $\mathcal{N}_\alpha$ for $\alpha=q^2B$. Since one can show that the  representations $\pi^B_{\theta,q}$ are $*$-isomorphisms onto their ranges (see \cite{DeN10,Fre13,ADT15}), this implies that the spectrum of  $H^B_{\theta,q}$ agrees with the spectrum of $\mathfrak{H}^{q^2B}$.  However, the latter is just the spectrum of $H^{q^2B}_{\rm Hof}$, i.e.\ it is again given by the black and white Hofstadter butterfly.

In order to associate Chern numbers with the spectral subbands of $H^B_{\theta,q}$, we now turn it by a suitable unitary transformation into matrix-multiplication form. Since $H^B_{\theta,q}$ contains within $\E^{\I q\mathcal{K}_2 }$ a shift by $q B$ in the $k_1$-direction, 
this is possible if we assume this shift to be a rational fraction of the width $\frac{2\pi}{q}$ of the Brillouin zone, i.e.\ $qB =  \frac{2\pi}{q} \frac{\tilde p}{\tilde q}$ or $B =   \frac{2\pi}{q^2} \frac{\tilde p}{\tilde q}$ with $\tilde p$ and $\tilde q$ coprime.
To this end we pass  from $\Hi_\theta$, i.e.\ from complex valued functions on the Brillouin zone $M^*_q=[0,\frac{2\pi}{q})\times [0,2\pi)$,  to $\C^{\tilde q}$-valued functions on the further reduced Brillouin zone  $M_{q,\tilde  q}^{*}=[0,\frac{2\pi}{q\tilde q})\times[0,2\pi)$.
To define the corresponding unitary map $U^B:\Hi_\theta \to L^2(M_{q,\tilde q}^*,\C^{\tilde q})$, we 
 let
\[
M_j := \left\{(k_1,k_2)\in M_q^*\,|\, k_1\in \left[  (j-1)qB, (j-1)qB+\tfrac{2\pi}{q\tilde q}\right)\right\}\;\mbox{ for }\; j=1,\ldots, \tilde q\,
\]
and define 
$$(U^B\psi)_{j}(k):=\E^{\I \theta k_2 (j-1) \frac{\tilde p}{\tilde q} }\,\psi\left(k_{1}+  (j-1)\,qB, k_{2}\right)\quad\mbox{ for }\quad k\in M_{q,\tilde q}^*\,.
$$
Thus $(U^B\psi)_{j}$ is obtained  by restricting $\psi\in \Hi_\theta$ to the region $M_j$, translating it to $M_{q,\tilde q}^*=M_1$ and finally multiplying it by $\E^{\I \theta k_2 (j-1) \frac{\tilde p}{\tilde q} }$. The last phase turns the translation by $qB$  in the $k_1$-direction on $\Hi_\theta$  into the cyclic permutation of components in $L^2(M_{q,\tilde q}^*,\C^{\tilde q})$ times a phase.
More precisely we have
\[
\E^{\I q \mathcal{K}_1 } \psi(k)=\E^{\I q k_{1}}\psi(k_{1},k_{2} ) \;\mbox{ and thus }\;
\left(U^B \,\E^{\I q \mathcal{K}_1 } \psi\right)_j(k)=\E^{\I q \left(k_{1}+ (j-1)qB\right)}\psi_j(k  )\,,
\]
and 
\[
\E^{\I q \mathcal{K}_2 } \psi(k)=\E^{\I qk_{2}}\psi(k_1+qB,k_2 ) \;\mbox{ and thus }\;
\left(U^B \,\E^{\I q \mathcal{K}_2 } \psi\right)_j(k)= \E^{\I  qk_{2}}\E^{-\I \theta k_2\frac{\tilde p}{\tilde q} }\psi_{j+1}(k)\,.
\]

Hence  $U^{B}H_{\theta,q}^{B}U^{B *}$ acts as
the matrix-valued multiplication operator
 \begin{equation}\label{Hthetaq}
H_{\theta,q}^{B}(k)=\footnotesize
\begin{pmatrix}
2 \cos (qk_{1}) &\E^{ \I  k_2 \big(q- \theta\frac{\tilde p}{\tilde q} \big) } & 0&\ldots&\E^{-\I  k_2 \big(q- \theta\frac{\tilde p}{\tilde q} \big) }\\[1mm]
\E^{-\I  k_2 \big(q- \theta\frac{\tilde p}{\tilde q} \big) } & 2 \cos (q(k_{1}+q B)) &\E^{ \I  k_2 \big(q- \theta\frac{\tilde p}{\tilde q} \big) }&\ldots&0\\[1mm]
0 &\E^{-\I  k_2 \big(q- \theta\frac{\tilde p}{\tilde q} \big) }& 2 \cos (q(k_{1}+2q  B)) &\ldots&0\\
\vdots& \ddots& \ddots&\ddots&0\\[1mm]
0&\ &\ &\ &\E^{\I  k_2 \big(q- \theta\frac{\tilde p}{\tilde q} \big) }\\[1mm]
\E^{ \I  k_2 \big(q- \theta\frac{\tilde p}{\tilde q} \big) }&0&\ldots&\E^{-\I  k_2 \big(q- \theta\frac{\tilde p}{\tilde q} \big) }&2 \cos (q(k_{1}+(\tilde q-1)q  B))
\end{pmatrix}\,.
\end{equation}

\begin{figure}
\floatbox[{\capbeside\thisfloatsetup{capbesideposition={right,center},capbesidewidth=5.2cm}}]{figure}[11cm]
{\caption{\small The operator $H^B_{-2,3}$ is up to a constant factor and higher order terms in the Fourier expansion of $E_2(k)$ the leading order part of the Peierls substitution Hamiltonian for the middle band of $H^{B_0}_{\rm Hof}$ for $B_0 = 2\pi\frac{1}{3}$. This band has Chern number $-2$. As can be seen from the coloring, the Chern numbers of the subbands of $H^B_{-2,3}$ match for $\frac{B}{2\pi}\in [0,\frac19]$ exactly the Chern numbers of the corresponding subbands of $H^{B_0+\tilde B}_{\rm Hof}$ with
$\tilde B = B \left( 1 - \frac{1}{1+\frac{ \pi}{3B}}
\right) = B(1+\Or(B))$. 
}\label{subbut}}
{\includegraphics[width=11cm]{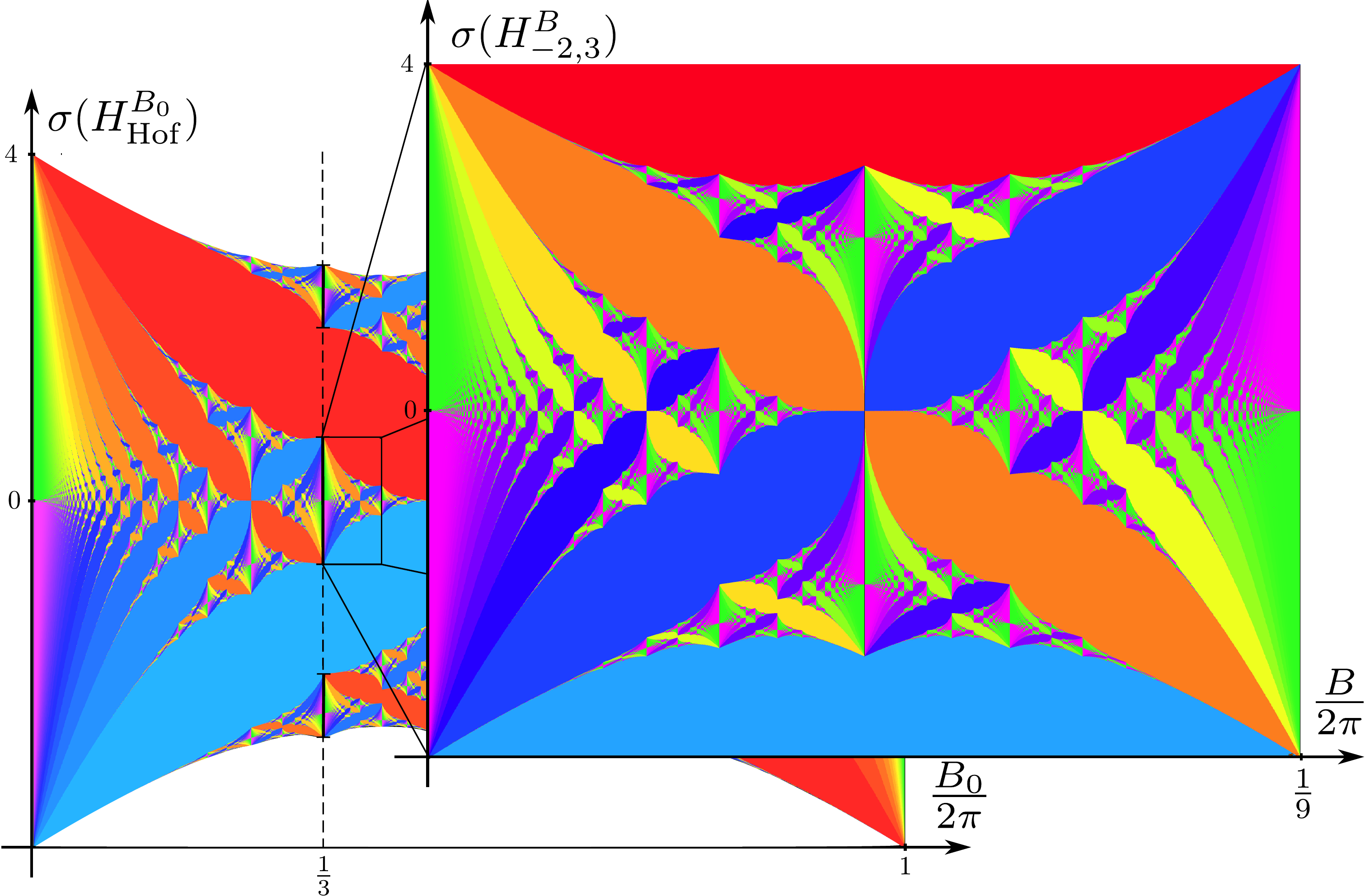}}
\end{figure}

Like the Hofstadter matrix $\hat H^{B_0}_{\rm Hof}(k)$, also $H_{\theta,q}^{B}(k)$ has $\tilde q$ distinct eigenvalue  bands 
$E_{\theta,q,n}^B(k)$, $n = 1,\ldots, \tilde q$. By isospectrality of $H_{\theta,q}^{B}$ and $H^{q^2 B}_{\rm Hof}$ the ranges of these band functions all agree. However, as functions they are in general distinct. The corresponding eigenprojections $P_{\theta,q,n}^B(k)$ define line bundles over the torus $M^*_{q,\tilde q}$ and
one can compute their Chern numbers  by   integrating the curvature of the corresponding Berry connection $P^B_{\theta,q,n}U^B\nabla^\theta_k U^{B*}$ over the reduced Brillouin zone $M^*_{q,\tilde q}$. Using a program from \cite{Amr15}, we did this numerically   for a large number of values for $\theta$, $q$ and $B$ and   found,  that the Chern numbers of the subbands of  $H_{\theta,q}^{B}(k)$ always match the Chern numbers of the corresponding sub-subbands of the Hofstadter Hamiltonian. To make this more precise, recall that  $H_{\theta,q}^{B}(k)$ was derived as the Peierls substitution Hamiltonian for a magnetic subband of $H^{B_0}_{\rm Hof}$ for $B_0 = 2\pi \frac{p}{q}$ with Chern number $\theta$ perturbed by a small additional magnetic field $B$. The Chern numbers of the subbands of $H_{\theta,q}^{B}(k)$ for $B = \frac{2\pi}{q^2}\frac{\tilde p}{\tilde q}$ agree with the Chern numbers of the subbands of $H^{B_0+\tilde B}_{\rm Hof}$ into which the unperturbed subband of $H^{B_0}_{\rm Hof}$ splits. Here  
\[
\tilde B =  B\,\left( 1 - \frac{1}{ 1-\frac{2\pi  }{ q\theta B} } \right)=  B\,\left( 1 - \frac{1}{ 1-\frac{q\tilde q  }{ \theta\tilde p } } \right)= B   +\Or(B^2) \,.
\]
The situation is depicted in Figure~\ref{subbut}. Note, however, that for drawing the colored butterfly of $H^B_{\theta,q}$ it is not feasible to compute all Chern numbers numerically by integrating the curvature of the Berry connection. This is because for large denominators $\tilde q$ the matrix $H^B_{\theta,q}(k)$ and the number of its subbands becomes large.
Instead in \cite{Amr15}   an algorithm was found, that allows to compute the  Chern numbers of $H^B_{\theta,q} $ in a purely algebraic fashion, similar to the diophantine equations used for labeling the gaps of $H^{B_0}_{\rm Hof}$.  
Also the code to produce the colored butterfly of $H^B_{-2,3}$ in Figure~\ref{subbut} is taken from \cite{Amr15} and based on a code originally developed by Daniel Osadchy.
This algorithm, the details on the numerics and a much more detailed study of the operator $H^B_{\theta,q}$ will be presented elsewhere \cite{ADT15}. There we also show how to explicitly incorporate  a better approximation to the true dispersion relation of a magnetic subband and    the subprincipal symbol as given in Theorem~\ref{5.11} into the  the Peierls substitution Hamiltonian. Then the  agreement in terms of Chern numbers depicted in Figure~\ref{subbut}  turns into a quantitative agreement also of the spectrum.
We take these numerical results  as an indication, that Theorem~\ref{5.11} also holds for perturbations by small constant magnetic fields.

\end{document}